\documentclass[useAMS,usenatbib]{mn2e}
\usepackage{graphicx}
\usepackage{natbib}
\usepackage{amsmath}
\title[Stable and unstable accretion in IM~Lup and RU~Lup]
{Stable and unstable accretion in the classical T~Tauri stars IM~Lup and RU~Lup 
as observed by {\it MOST\/}}

\author[M. Siwak et al.]
{Michal Siwak$^{1}$\thanks{E-mail: siwak@astro.as.up.krakow.pl},
Waldemar Ogloza$^{1}$,
Slavek M.\ Rucinski$^2$,
Anthony F.\ J.\ Moffat$^3$,\newauthor
Jaymie M.\ Matthews$^4$,
Chris Cameron$^5$,
David B.\ Guenther$^6$,
Rainer Kuschnig$^{4,9}$,\newauthor
Jason F.\ Rowe$^7$,
Dimitar Sasselov$^8$,
Werner W.\ Weiss$^9$\\
$^1$Mount Suhora Astronomical Observatory, Cracov Pedagogical University,
ul.\ Podchorazych 2, 30-084 Cracov, Poland\\
$^2$Department of Astronomy and Astrophysics,
University of Toronto, 50 St.\ George St., Toronto,
Ontario, M5S~3H4, Canada\\
$^3$D\'{e}partment de Physique, Universit\'{e} 
de Montr\'{e}al, C.P.6128, Succursale: Centre-Ville,
Montr\'{e}al, QC, H3C~3J7, Canada\\
$^4$Department of Physics \& Astronomy, University of
British Columbia, 6224 Agricultural Road, Vancouver, B.C., V6T~1Z1, Canada\\
$^5$Department of Mathematics, Physics \& Geology, Cape Breton University, 
1250 Grand Lake Road, Sydney,NS, B1P 6L2, Canada\\
$^6$Institute for Computational Astrophysics,
Department of Astronomy and Physics,
Saint Marys University, Halifax, N.S., B3H~3C3, \\
Canada\\
$^7$NASA Ames Research Center, Moffett Field, CA 94035, USA\\ 
$^8$Harvard-Smithsonian Center for Astrophysics,
60 Garden Street, Cambridge, MA 02138, USA\\
$^9$Universit\"{a}t Wien, Institut f\"{u}r Astrophysik, 
T\"{u}rkenschanzstrasse 17, A-1180 Wien, Austria\\
}

\date{Accepted 2015 December 2;  Received 2015 November 11; in original form 2015
September 6}

\pubyear{2015}

\begin{document}

\label{firstpage}

\maketitle

\begin{abstract}
Results of the time variability monitoring of the two classical 
T~Tauri stars, RU~Lup and IM~Lup, are presented. 
Three photometric data sets were utilised: 
(1)~simultaneous (same field) 
{\it MOST\/} satellite observations over four weeks in each of the
years 2012 and 2013, (2)~multicolour observations at the 
{\it SAAO} in April -- May of 2013, (3)~archival $V$-filter
{\it ASAS} data for nine seasons, 2001 -- 2009. They were
augmented by an analysis of 
high-resolution, public-domain {\it VLT-UT2 UVES} spectra 
from the years 2000 to 2012. 
From the {\it MOST\/} observations, we infer that irregular light 
variations of RU~Lup are caused by stochastic variability of hot 
spots induced by unstable accretion. 
In contrast, the {\it MOST\/} light curves of IM~Lup are fairly regular 
and modulated with a period of about 7.19 -- 7.58~d, which is in accord 
with {\it ASAS} observations showing a well defined $7.247\pm0.026$~d 
periodicity.
We propose that this is the rotational period of IM~Lup 
and is due to the changing visibility of two antipodal hot 
spots created near the stellar magnetic poles during 
the stable process of accretion.
Re-analysis of RU~Lup high-resolution spectra 
with the Broadening Function approach reveals signs of a large polar 
cold spot, which is fairly stable over 13 years. 
As the star rotates, the spot-induced depression of 
intensity in the Broadening Function profiles 
changes cyclically with period 3.71058~d, which was previously found by 
the spectral cross-correlation method. 
\end{abstract}


\section{Introduction}
\label{intro}

Continuous, high-precision, photometric observations obtained with 
{\it Microvariability and Oscillation of STars (MOST)\/} and {\it COnvection ROtation and planetary Transits (CoRoT)\/} 
space telescopes significantly contributed to the knowledge of a variety 
of mechanisms causing stellar variability of T~Tauri stars occurring 
on time scales of weeks and days. 
In particular, our first, long, nearly uninterrupted {\it MOST\/} 
light curves of the classical T~Tauri-type star
(CTTS) TW~Hya \citep{ruc08, siwak11a} and the {\it CoRoT} light 
curves of several CTTSs - members of the young cluster NGC~2264
\citep{alencar10},
permitted the first observational tests of 
theoretical three-dimensional magneto-hydrodynamical (3D MHD) models 
of plasma accretion on a magnetised star from the inner accretion 
discs \citep{kurosawa13}. 
The results of these numerical simulations are presented in detail 
in the series of papers by \citet{romanowa04, romanowa08, kulkarni08, 
romanowa09, kulkarni09}.

According to the authors of these theoretical investigations,
for magnetospheres a few times the stellar radius in size, 
the accretion from the surrounding disc can occur in a stable, 
a moderately stable or an unstable regime. 
The regime of accretion which dominates at a given time is determined 
by several factors: the magnetic field strength, the star's rotation rate, 
the misalignment angle $\theta$ between the stellar rotational axis and 
the magnetic poles; however, the dominant factor is the mass accretion rate \citep{kulkarni08}.
For a low value of the accretion rate, stable 
accretion takes the form of steady plasma flows from the inner disc toward 
the stellar magnetic poles in two funnels encircling the magnetosphere \citep{romanowa04}.
The funnels produce two antipodal, banana-shaped 
hot spots which are almost stationary on the star. 
The resulting light curves of such a CTTS are modulated with the 
rotational frequency of the star only.

An increased mass-accretion rate may lead to substantial diversification 
of plasma density in the inner accretion disc and to an onset 
of Rayleigh-Taylor (RT) instabilities. 
The instabilities produce a few equatorial tongues, in which 
the plasma is first transferred directly from the disc toward 
the star, but ultimately it is wound along stellar magnetic 
field lines, which form miniature funnel flows. 
The plasma impacts the star at moderate latitudes (35-65~deg) 
and produces hot spots which are smaller than those produced by 
the stable accretion funnels \citep{romanowa08, kulkarni08, kulkarni09}. 
The chaotic behaviour of the tongues and the resulting small 
hot spots lead to progressively more chaotic  
light curves as more spots are formed: While at 
the beginning of the RT instabilities the funnel component still produces 
a single peak at the stellar rotation frequency (though not 
as steady as for the purely stable case), 
as time progresses the frequency spectrum of the
star brightness starts to show increasingly 
more additional peaks produced by the complex hot-spot distribution
around the star.
This stage is called a moderately stable accretion regime.

In the fully unstable regime, for mass accretion rates higher 
by more than an order of magnitude than in the stable regime, the hot 
spots are created by only a few unstable tongues rotating around the star 
with angular velocity of the inner disc. 
As the shape, intensity, number and position of the hot spots 
change on the inner-disc dynamical timescale, 
the resulting phenomena are highly chaotic and no stellar-rotation
frequency is detectable in complex light curves or their 
Fourier or wavelet spectra. 

The type of accretion may alternate between the different regimes 
depending on changes in the mass accretion rate \citep{romanowa08}. 
This was likely observed in TW~Hya, where the accretion 
regime was found to switch between a moderately stable and 
an unstable regime \citep{siwak14}. 
A similar effect was also found in the classical T~Tau star
 2MASS~06401258+1005404 observed 
twice by {\it CoRoT}; in 2008 the star showed a light-curve modulated with 
a stable spot, while in 2011 the light curve was irregular and dominated 
by accretion bursts \citep{stauffer14}. 
Prompted by the most recent results obtained for TW~Hya \citep{siwak14}, 
both the probable discovery of the accretion-regime alternation 
and the discovery of short-term occultations, we used
 {\it MOST\/} in a continuation of our previous, already cited
work, but within a broader context of investigation of young stars 
(\citealt{ruc10, siwak11b, siwak13, cody13, windemuth13, zwintz13}). 
For the present analysis, we have chosen two well-known southern stars 
RU~Lup and IM~Lup, the brightest among the six known CTTSs in the
Lupus~2 Star Forming Region \citep{comeron08}. 
Both stars could be observed simultaneously in the same {\it MOST\/} field.

We describe previous results obtained for our targets in 
Section~\ref{history}. 
The high-cadence data for RU~Lup and IM~Lup obtained with 
the {\it MOST\/} satellite during 4 weeks 
in 2012 and 2013 are described in Section~\ref{observations}. 
In the same section, we describe supplementary multicolour 
observations of these stars obtained in 2013
at the South African Astronomical Observatory ({\it SAAO}) and
 long-term, 2001--2009 All Sky Automated Survey ({\it ASAS}) data. 
The  spectroscopic data obtained 
during 2000, 2002, 2005, 2008 and 2012 in the public-domain at the European Southern Observatory 
({\it ESO}) using the {\it La Silla 3.6~m} 
and {\it Cerro Paranal VLT-UT2} telescopes 
are described in Section~\ref{spectra}.
We summarise and discuss the results obtained in the context of 
recent theoretical investigations of accretion flow 
in magnetised stars in Section~\ref{discussion}.

\section{History and previous observations}
\label{history}

\subsection{RU~Lup}
\label{rulup}

The major peculiarity of the star with the spectrum dominated 
by strong and broad emission lines was first noticed by 
\citet{Merrill41} and \citet{joy45}. 
The emission lines of RU~Lup belong the broadest among CTTSs:
The H$\alpha$ emission 
line equivalent width $EW(H\alpha)$ is always larger 
than 100~\AA, with maximum value as large as 216~\AA\
\citep{appenzeller83}.
\citet{gahm74, gahm79} concluded that the emission lines and 
the severe veiling of the absorption spectrum
(first detected by \citealt{schwartz81}) are likely produced by a hot 
and very bright source.

Large and irregular light variations of RU~Lup have been observed
over long time scales in Harvard photographic 
plate data obtained between 1893 and 1912 showing  
variability in the range of 9--11~mag \citep{joy45}. 
This was confirmed by numerous observations by amateur astronomers 
in the twentieth Century, who observed variability 
in the similar range of 8.6--12.8 mag \citep{lamzin96}.
Although the first photoelectric observations of \cite{Hoffmeister58},
as analysed by \cite{Plagemann69}, indicated a quasi-periodic oscillation 
at a period of $P=3.7\pm0.2$~d, their subsequent analysis of \cite{giovannelli91} 
questioned the statistical significance of this detection. 
The authors were also unable to find any periodicity in the 
second data-set of RU~Lup observations 
published by \citet{Hoffmeister65}. 

Irregular variability of this star was confirmed 
with more accurate broad-band observations in the whole
optical -- IR spectral range of photometric
bands between $U$ and $M$.
As firmly stated by \citet{gahm74}, \citet{hutchinson89} and 
\citet{giovannelli91}, the greatest amplitude 
of light variations is observed in the $U$-filter, its value 
slowly decreasing towards longer wavelengths, 
with the star being redder (or better said -- less blue) when fainter.
In accordance with findings from spectroscopy, the authors 
proposed that this could be due to rotation of the 
spotted stellar surface containing a hot source with an
estimated temperature $T$ of $\simeq 12000$~K. 
 
Once the model of magnetically-controlled accretion 
was formulated for CTTS \citep{konigl91} and successfully 
applied for interpretation of new multi-wavelength data 
of RU~Lup by \citet{giovannelli95} and \citet{errico96}, 
\citet{lamzin96} proposed that the observed variability 
and its spectral properties could be produced by 
a hot, ring-like accretion belt (6500~K) formed around 
the two magnetic poles on the star. 
The ring could occupy about 30\% of the stellar surface.
The hot-spot properties were later substantially modified 
by \citet{stempels02}: Based on the 
veiling factors $vf$
\footnote{The spectral veiling factor 
$vf$ measures changes in depth of absorption lines
due to some, frequently unexplained, emission component: 
$vf=|EW_{\rm unveiled}|/|EW_{\rm T~Tau}|-1$, 
where $EW_{\rm unveiled}$ and $EW_{\rm T~Tau}$ 
are the absorption-line equivalent widths 
of an unveiled reference spectrum and a veiled T~Tauri star, 
respectively.}, which decrease monotonically from $vf \simeq 15-20$ at 4091~\AA\ to 
$4-5$ at 6210~\AA, the authors proposed a hot spot 
with $T\simeq 10000$~K, but with a smaller 3-6\% surface filling factor.

\citet{lamzin96} found that the majority of the observed luminosity of
the star is produced by accretion processes and 
only 10\% by the star itself. This circumstance leads 
to difficulties in determination of RU~Lup's stellar parameters. 
The authors estimated the stellar parameters as follows: 
the effective temperature $T_{\rm eff}=3800$~K (Sp M0.5), 
the luminosity $L=0.49$~L$_{\odot}$ 
and the radius $R=1.62$~R$_{\odot}$. 
From its position on the Hertzsprung-Russell diagram they estimated the 
mass to be smaller than a solar mass, the age at about 1~Myr 
and the mass accretion rate at $3 \times 10^{-7}$~M$_{\odot}$~yr$^{-1}$. 
\citet{herczeg05} proposed a slightly lower rate of 
$5\pm2 \times 10^{-8}$~M$_{\odot}$~yr$^{-1}$. 
Based on their estimate of the interstellar extinction of $A_V=0.07$, distance 140~pc,
and veiling factors given by \citet{stempels02}, 
they modified the stellar parameters to
$L=0.6$~L$_{\odot}$ and  $R=1.64$~R$_{\odot}$. 
The age remained similar as before (2--3~Myr), while the mass was 
lowered to 0.6-0.7~M$_{\odot}$.
\citet{stempels02} re-determined some parameters using the spectral region 
5905--5965~\AA\ where the spectrum shows no obvious 
emission components associated with metallic lines, as follows: 
$T_{\rm eff}=3950$~K, 
surface gravity $\log g=3.9$, 
the mean radial velocity $v_{\rm rad}=-1.9\pm0.02$~km/s, 
and spectral lines rotational broadening $v\,\sin i=9.1\pm0.9$~km/s, 
where $i$ is the inclination angle between the stellar rotational axis 
and the observer line of sight.
They concluded that such a small value of the spectral 
line broadening, as well as the presence of emission lines 
containing blueshifted signatures of the collimated outflow, 
indicate that 
RU~Lup is observed at a low inclination angle\footnote{This 
makes the star similar to TW~Hya, which shows very strong
accretion processes between the inner disc and the star and has been
extensively observed by us using {\it MOST\/}.}.

Although no obvious periodic signal is present in any photometric data, 
\citet{stempels07} found a period of $3.71058\pm0.0004$~d 
using the cross-correlation function analysis of the
radial velocity variations. 
The periodicity appeared to be stable over many years.
Assuming that this is the rotational period of the star, 
and that the stellar radius is 1.64~R$_{\odot}$, 
the mass 0.65~M$_{\odot}$ and 
the observed $v\,\sin i = 9.1 \pm 0.9$~km/s, 
they determined the inclination of RU~Lup at 24~deg. 
In order to explain the observed amplitude in radial 
velocity changes (2.2~km/s), which is anti-correlated 
with the bi-sector of the absorption lines, the authors concluded 
that RU~Lup must be covered by relatively large, cold spots. 
One possibility is a single spot with a radius 
of more than 30~deg, an average temperatures in the range 
3000-3600~K and a magnetic field strength of 3kG. 
The authors did not find any short-term correlation between 
the measured radial velocities 
and the simultaneous, moderate-quality photometric {\it ASAS} data. 
In the following sections of the paper it becomes obvious that the 
expected signal variations from the large cold 
spot (0.15~mag, at most) is totally suppressed by irregular variations 
of much larger amplitude due to unstable accretion.

\begin{figure*}
\includegraphics[height=60mm,angle=0]{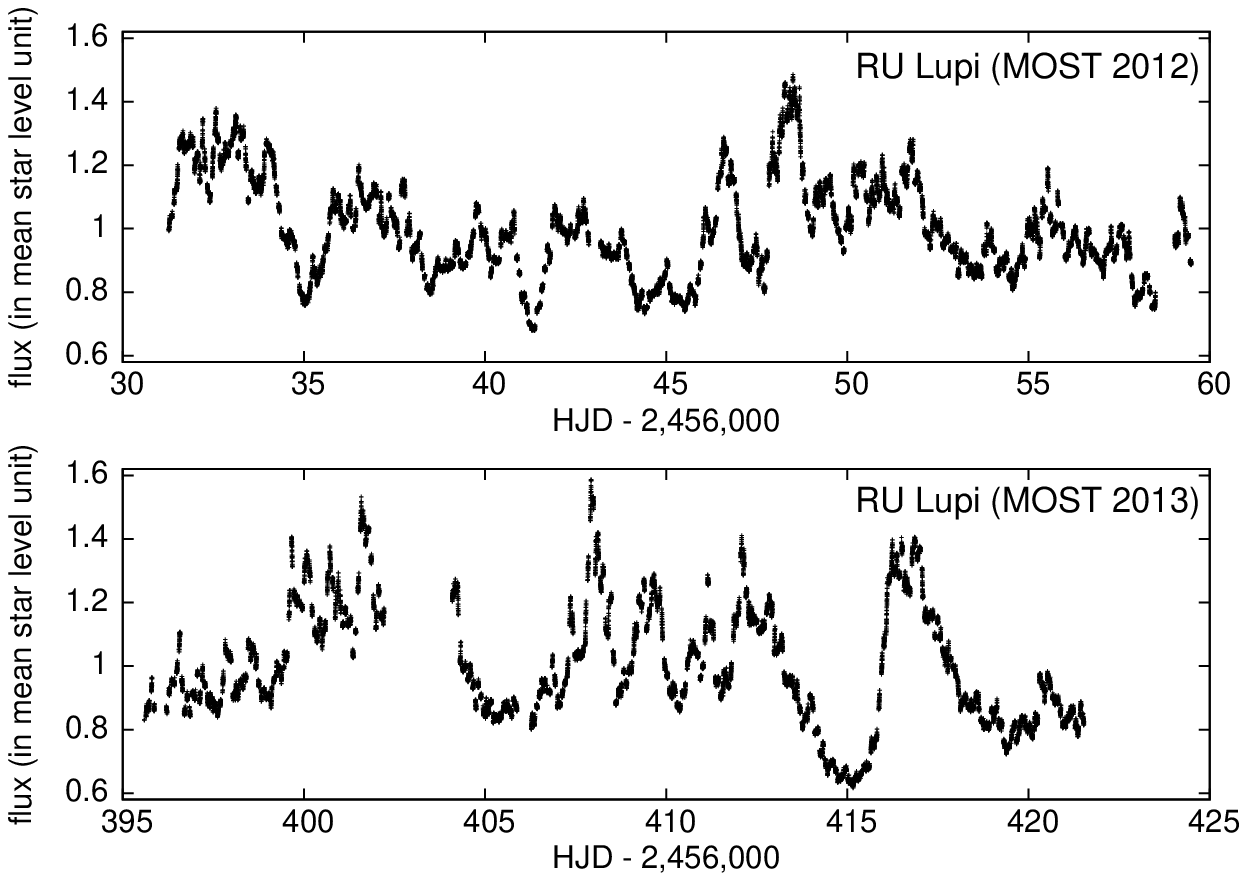}
\includegraphics[height=60mm,angle=0]{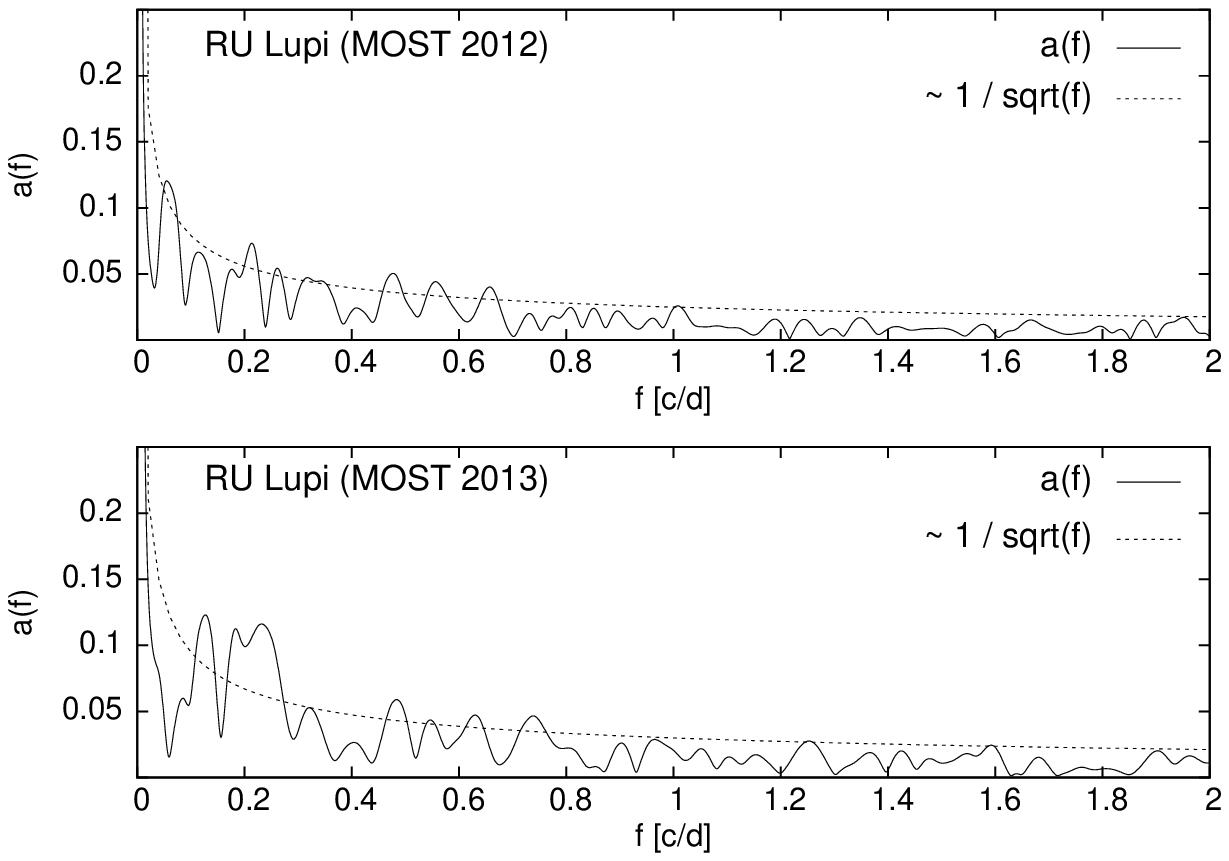}
\includegraphics[height=60mm,angle=0]{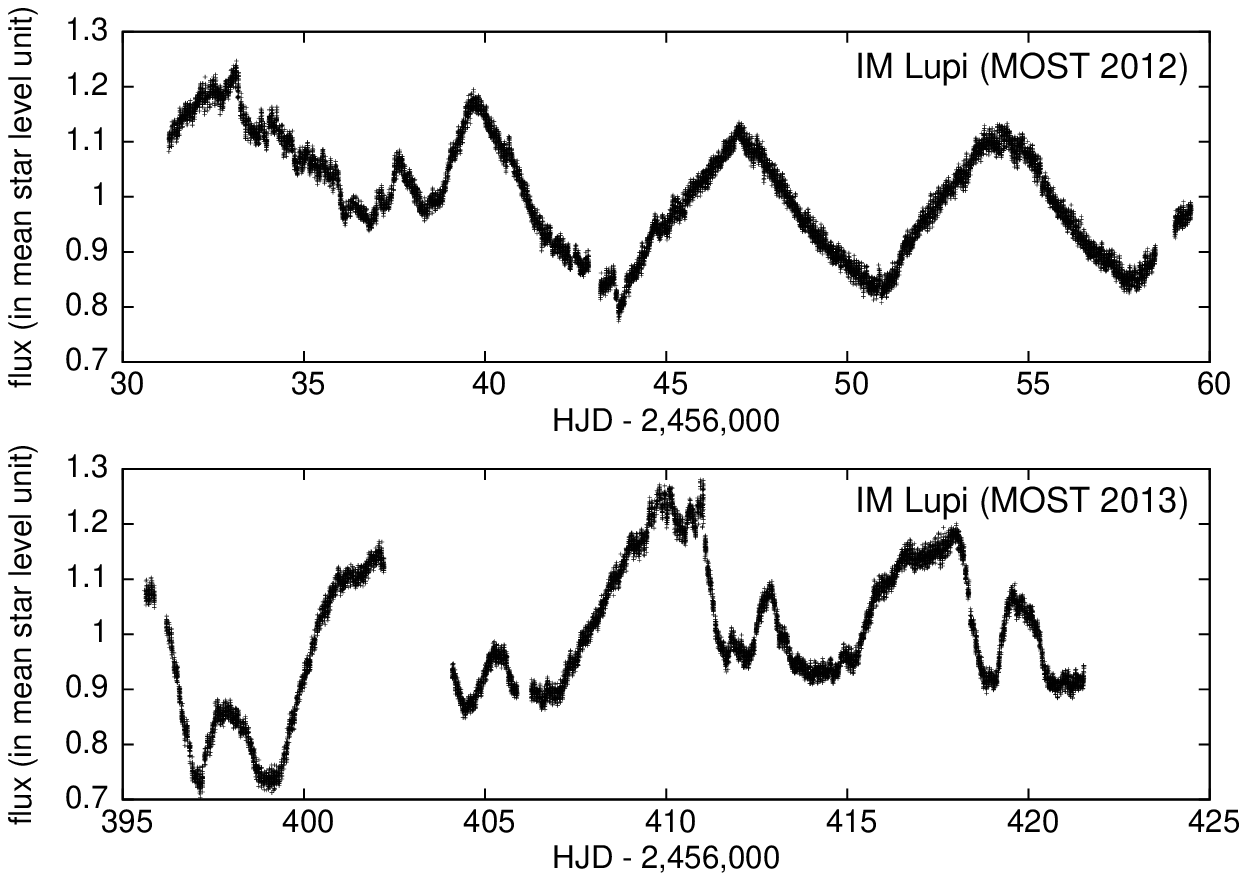}
\includegraphics[height=60mm,angle=0]{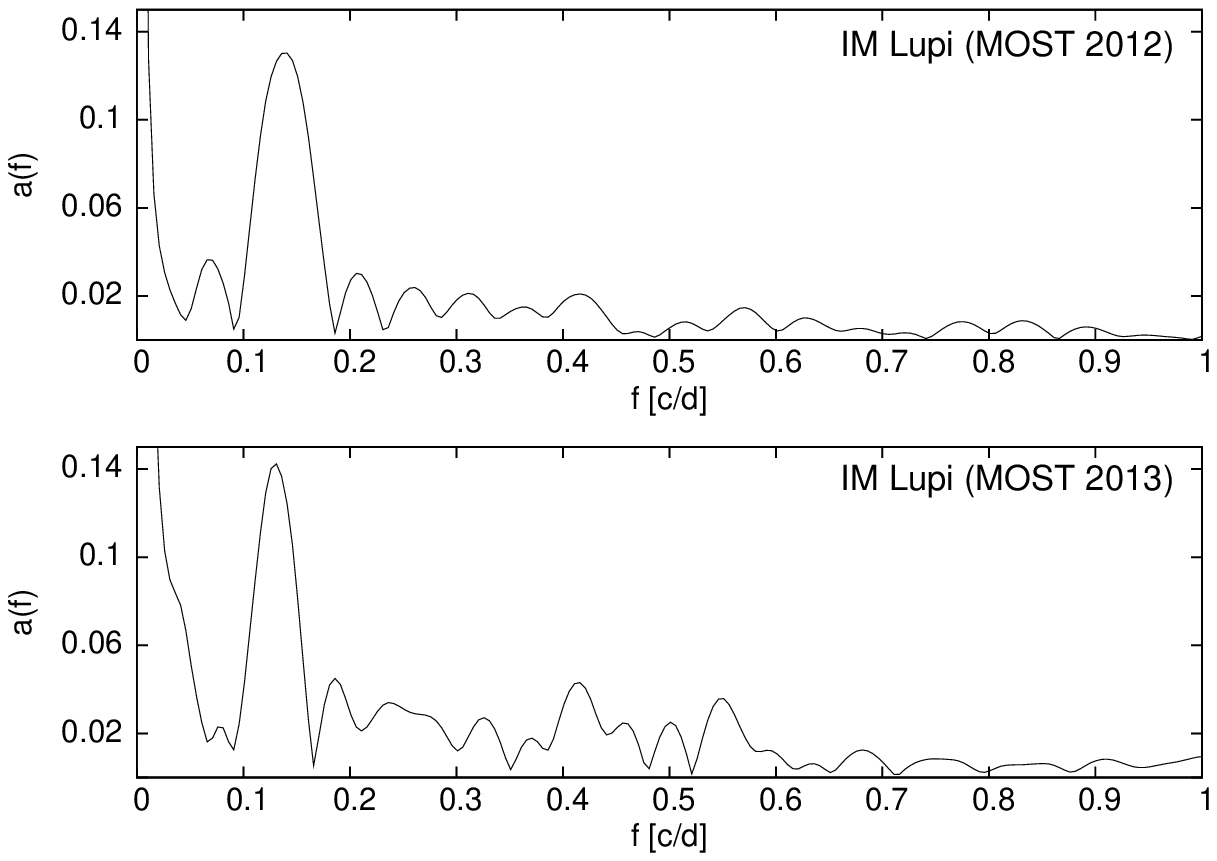}
\caption{The left panels show the observed light variations of RU~Lup and IM~Lup obtained 
by {\it MOST\/} during 2012 and 2013; they are expressed in Heliocentric Julian Days ($HJD$) 
and in flux units scaled to unity at the mean stellar brightness level. 
The right panels show the corresponding amplitude frequency spectra $a(f)$, 
where $f$ is the frequency in cycles per day. 
The amplitude errors in the spectra are negligible, in practice coincident with 
the X-axis, therefore omitted for clarity. 
For RU~Lupi, the frequency spectra show also an arbitrarily scaled flickering noise 
envelope with $a(f) \propto 1/\sqrt f$.
}
\label{Fig.dat}
\end{figure*}

\subsection{IM~Lup (Sz~82)}
\label{imlup}

In contrast to RU~Lup, IM~Lup shows a typical absorption-line 
spectrum free of any obvious peculiarities. 
In spite of the young age of 1~Myr, 1.8~Myr or 2.5~Myr, as respectively 
derived by \citet{finkenzeller87}, 
\citet{batalha93} and \citet{martin94}, 
lack of any optical veiling and rather weak emission lines 
prompted the previous investigators 
to classify IM~Lup as a chromospherically 
active, weak-lined T~Tauri-type Star (WTTS).
The hydrogen emission is relatively weak, with $EW(H\alpha)$ 
occasionally as small as 5.2~\AA~\citep{batalha93}, 
4.7~\AA~\citep{reipurth96}, 4~\AA~\citep{wichmann99} 
or even 2.8--5.2~\AA~\citep{gunther10}.
These values are significantly smaller than the traditional 
limit of 10~\AA~defining the borderline between CTTS and WTTS. 
Yet, there is strong evidence that the above classification is 
incorrect and the star belongs to the class of classical T~Tauri stars.
\citet{batalha93} observed the hydrogen line emission 
of IM~Lup at a much higher level with $EW(H\alpha)$ staying for 
two years at an average level of 17~\AA~and varying within 
the range of 7.5--21.5~\AA. 
Considerable asymmetry of the $H\alpha$ profiles in IM~Lup 
\citep{reipurth96, gunther10, valenti03} supports the view of an 
ongoing accretion and thus its CTTS membership. 
As stated by \citet{gunther10}, WTTS are generally expected
 not to show accretion, but for many CTTS the $EW(H\alpha)$ 
is variable, suggesting a low or variable mass-accretion rate. 
The authors estimate the mass accretion rate in IM~Lup 
at only $10^{-11}$~M$_{\odot}$~yr$^{-1}$.

Other evidence of the CTTS membership of IM~Lup came 
with the discovery of a large (400--700~AU), gas-rich 
accretion disc \citep{vanKempen07}. 
Based on {\it Hubble Space Telescope} high-contrast images, \cite{pinte08} 
determined its inclination at $50\pm5$~deg. 
They confirmed that the disc is large, 400~AU across, for the 
{\it Hipparcos's\/} spacecraft derived distance 
$d=190\pm27$~pc (as given by \citealt{wichmann98}) and 
massive, 0.1~M$_{\odot}$. 
From spectral energy distribution modelling, 
they estimated the value of the inner disc radius at 0.15~AU. 
Recent sub-mm observations of \citet{panic09} show that the disc 
is in Keplerian rotation around a central 
mass of $1.2\pm0.4$~M$_{\odot}$. 
The authors discovered a break in the
surface density of the disc, which initiated space- 
and ground-based searches for a Jupiter-like planet 
around the star \citep{mawet12}.

The first set of stellar parameters of IM~Lup, as derived by 
\citet{batalha93}, was recently refined by \citet{pinte08}.
With the assumption of a distance of $190\pm27$~pc 
\citep{wichmann98} they obtained $T_{\rm eff}=3900$~K 
(Sp.\ M0V), $\log g=3.5$, $R=3$~R$_{\odot}$, $L=1.9$~L$_{\odot}$, 
$A_V=0.5$~mag and age of 1~Myr.
From the high-resolution {\it HARPS} spectra \citet{gunther10} 
obtained $v\,\sin i=15\pm2$~km/s. 
Using $R=3$~R$_{\odot}$ and $i=50\pm5$~deg of \citet{pinte08} 
or $i=54\pm3$~deg of \citet{panic09}, and assuming that the accretion disc 
is coplanar with the stellar equator, one can estimate possible values 
of the rotational period at $7.7\pm0.7$ and $8.1\pm0.4$~d, respectively. 
Interestingly \citet{batalha98} found the signature of a 
7.42~d period in archival photometric data, although
the significance of this periodicity was very low and the authors 
concluded that IM~Lup is an irregularly variable star.

\section{Photometric observations}
\label{observations}


\subsection{MOST observations}
\label{mostobs}

The {\it MOST\/} satellite has been described several times
in our previous papers of this series, as listed in
Section~\ref{intro}.
The pre-launch characteristics of the mission are described by 
\citet{WM2003} and the initial post-launch performance by \citet{M2004}.
The satellite observes in one broad-band filter covering 
the spectral range of 350 -- 700~nm with effective 
wavelength similar to that of the Johnson $V$-filter.

The {\it MOST\/} observations of RU~Lup and IM~Lup were done
simultaneously, with both stars in the same field of the satellite.
The first series of observations, conducted in  
{\it direct-imaging\/} data-acquisition mode, 
took place over 28.216 days between 13 April and 12 May, 2012, 
during 368 satellite orbits. 
Encouraged by the first results, we re-observed the stars 
during the next observing season between 13 April and 8 May, 2013, 
over 25.932 days during 334 satellite orbits.  
The stars are not in the Continuous Visibility Zone 
of the satellite so that their observations were
interrupted for 40-70~min during each satellite orbit 
by observations of secondary targets, 
and occasionally for $\sim1$~d by time-critical observations 
of other targets. 
As a result, 
the effective total time coverage for our targets 
during the 2012 and 2013 runs
was 12.043 days (42.7\%) and 8.331 days (32.1\%), respectively.

The individual exposures were always 60~s long. For photometric 
reductions, the {\it dark\/} and {\it flat\/} calibration frames 
were obtained by averaging 
a dozen empty-field images specifically taken during each run.
Aperture photometry was carried out on the {\it dark-\/} and {\it flat-\/} 
corrected images by means of the {\small \sc DAOPHOT~II} 
package \citep{stet}.
As in our previous investigations, weak correlations between 
the stellar flux, its position on the chip, and the sky 
background level within each {\it MOST\/} orbit were removed. 
We obtained a well-defined light curves expressed in Heliocentric 
Julian Days $HJD$ and normalized flux units (Figure~\ref{Fig.dat}, 
the left side panels) for almost the whole duration of the observations, with typical error 
(i.e. the standard deviation) of a single data point of 0.010--0.011~mag (RU~Lup) 
and 0.016--0.017~mag (IM~Lup) in the 2012 and 2013 runs, respectively. 
In addition to the targets, the nearby bright GCS~07838-00926 companion, 
separated by 20~arcsec from IM~Lup was also  
analysed in the same way. 
The star remained constant during 2012 and 2013.

The light variations of RU~Lup in the {\it MOST} data appear to be irregular 
with no sign of any periodic features, either in the 2012 or in the 2013 
light curves. 
This behaviour is typical for Type~II, irregular and accretion-burst classes, 
as defined by \citet{herbst94}, \citet{alencar10} and \citet{stauffer14}, respectively. 
In contrast, the light curves of IM~Lup show a fairly stable 
7--8~d periodicity, which is typical for Type~IIp or spot-like CTTSs, as defined 
in \citet{herbst94} and \citet{alencar10} classification schemes. 
Only a part of the 2012 light curve of IM~Lup is disturbed by small-scale, 
irregular variations, which disappeared after $HJD \approx 2,456,038$. 
Interestingly, while the 2012 light curve shows regularly 
repeating maxima, the 2013 one shows a sequence of alternating  
higher and smaller brightness increases whose shape evolves slightly 
during each double-period sequence. 
We note that the second light curve is very similar 
to the synthetic curve obtained for the case 
of a stable accretion regime by \citet{kurosawa13} (see in Fig.~5 
of their paper).
In addition, we also carefully examined the light curves of
RU~Lup and IM~Lup for the 
presence of short brightness dips like those observed for 
TW~Hya \citep{siwak14}, with negative result.


\subsubsection{Frequency analysis of the {\it MOST} data}
\label{MOSTfr}

We performed frequency analysis of the light curves in a similar way
as in our previous papers of this series (see Section~4.2 in \citealt{ruc08}).
The bootstrap sampling technique permitted evaluation of 
the mean standard errors of the amplitude $a(f)$, 
where $f$ is the frequency (Fig.~\ref{Fig.dat}, the right panels).

We used all data points gathered by {\it MOST\/} 
in the Fourier spectrum calculations for RU~Lup. 
For IM~Lup, we removed the first 8 days of the 2012 data due 
to highly irregular light variations 
which appear to spoil the primary signal 
in the Fourier spectrum. 
Also the long-term ascending trend visible in the 2013 IM~Lup light 
curve was removed by a simple parabolic fit, 
as it had measurable impact on the amplitude and the 
period obtained from the frequency spectrum. 

The 2012 and 2013 Fourier spectra of RU~Lup show 
dominant flicker-noise characteristics, 
i.e.\ the amplitudes scale as $a(f) \propto 1/\sqrt f$, exactly 
as for TW~Hya \citep{ruc08,siwak11a, siwak14}, with no single 
periodicity dominating in the frequency spectra.
However, the corresponding spectra of IM~Lup clearly show a single, 
very persistent signal at about 7.5~d 
and with amplitudes similar in both years 
(Figure~\ref{Fig.dat}, right panels). 
Measurements of the central positions of the peaks 
result in $f=0.139\pm0.023$~c/d ($P=7.19$~d) 
and $f=0.132\pm0.020$~c/d ($P=7.58$~d) for the 2012 
and 2013 signals, respectively.
These values are in good agreement 
with the rotation period of 
the star estimated from analysis of the
spectroscopic, high-contrast imaging 
and interferometric observations 
(see the last paragraph of Section~\ref{imlup}).

\subsubsection{Wavelet analysis of the {\it MOST} data}
\label{wavmost}

As demonstrated during TW~Hya light-curve analyses 
by \citet{ruc08} and \citet{siwak11a}, 
 wavelets are a very useful tool for finding short-lasting
regularities in apparently irregular light curves of CTTS. 
While Fourier spectra of TW~Hya's 2008 and 2009 data 
taken over whole observing runs showed only flicker-noise 
characteristics, the wavelets revealed the presence of short 
oscillatory behaviour with quasi-periods which tended to progressively
decrease with time. Thus, wavelets permitted a simple and
direct method to follow up the time period progression versus time.

\begin{figure*}
\includegraphics[width=85mm]{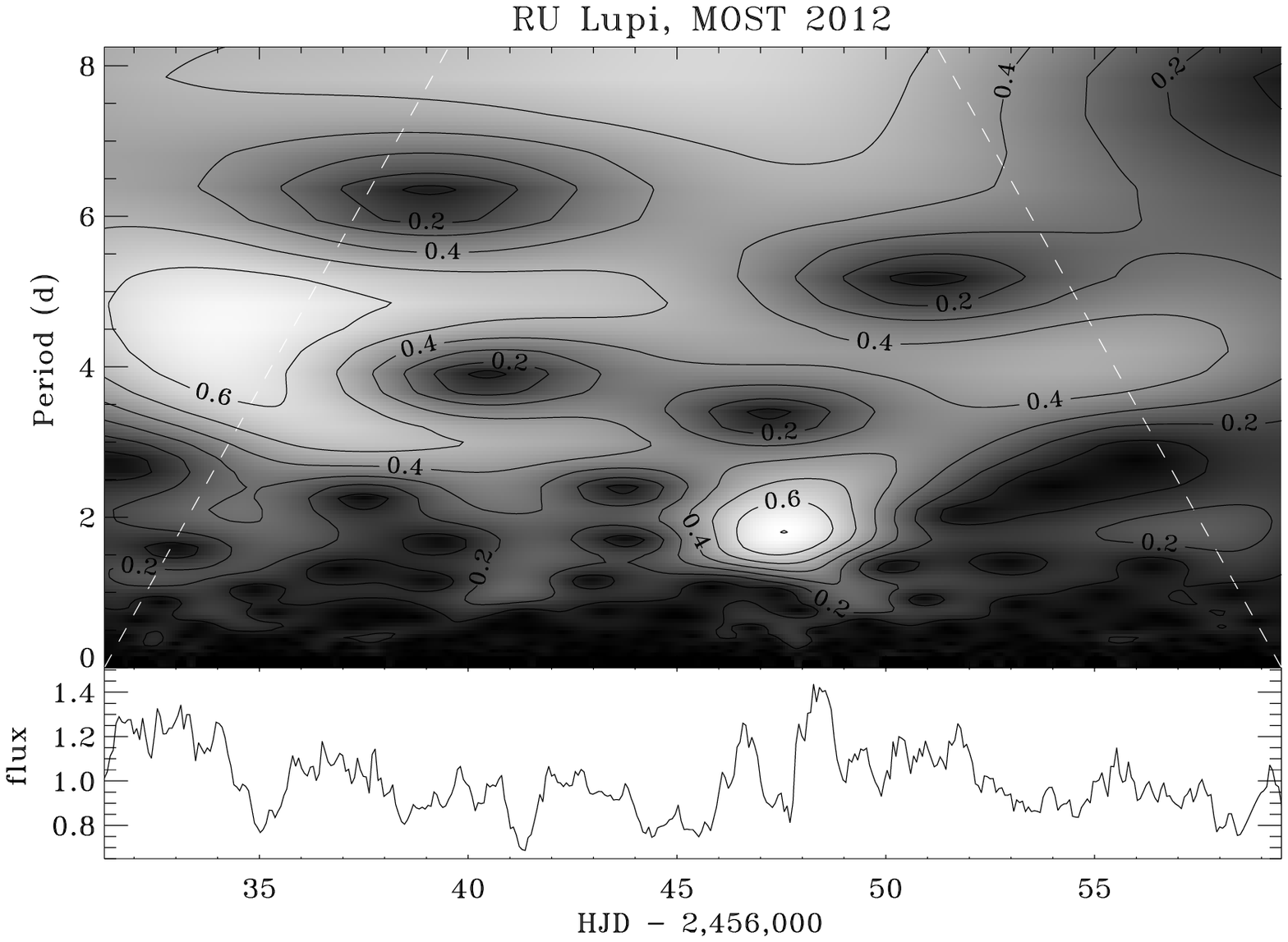}
\includegraphics[width=85mm]{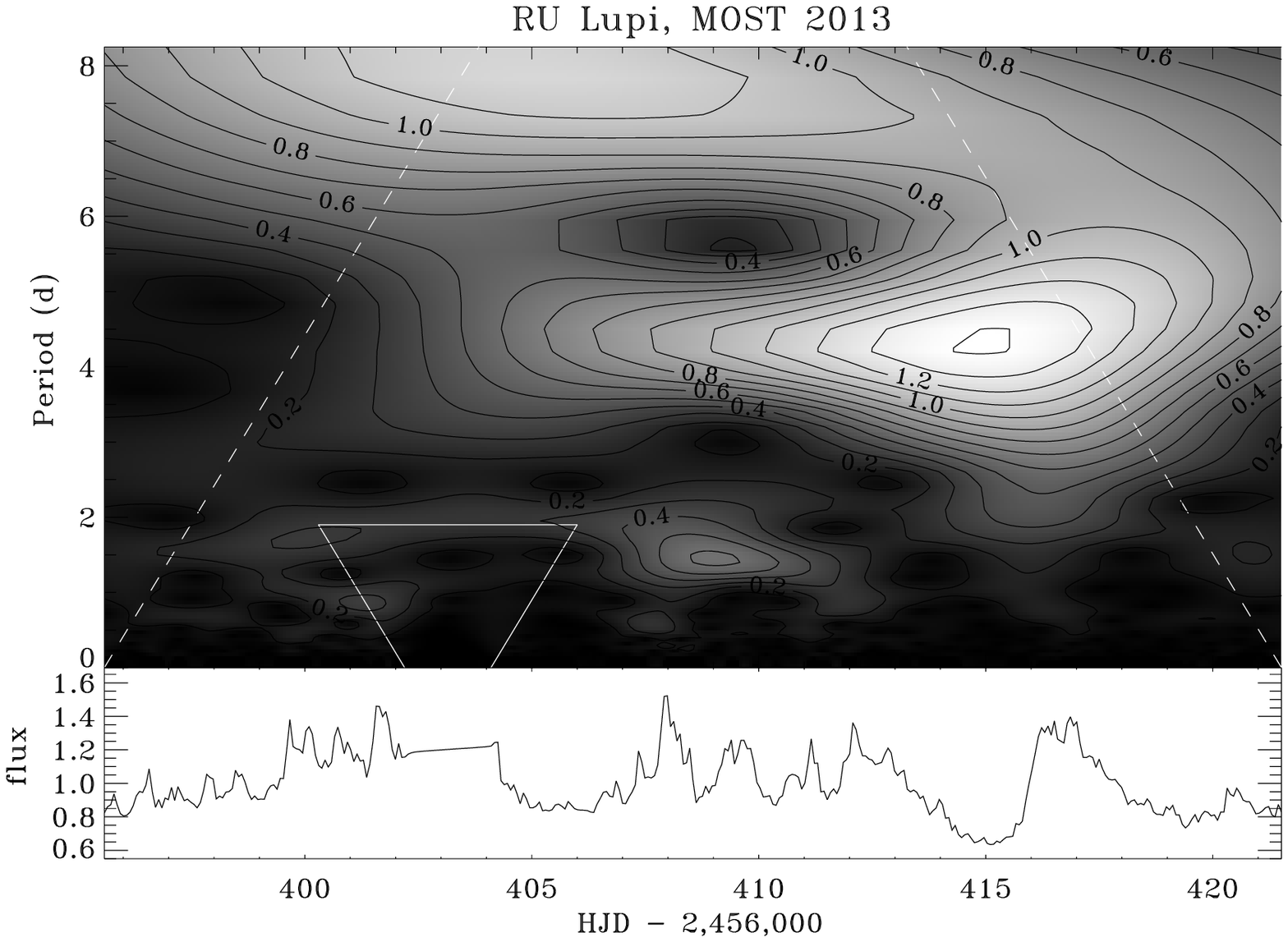}\\
\includegraphics[width=85mm]{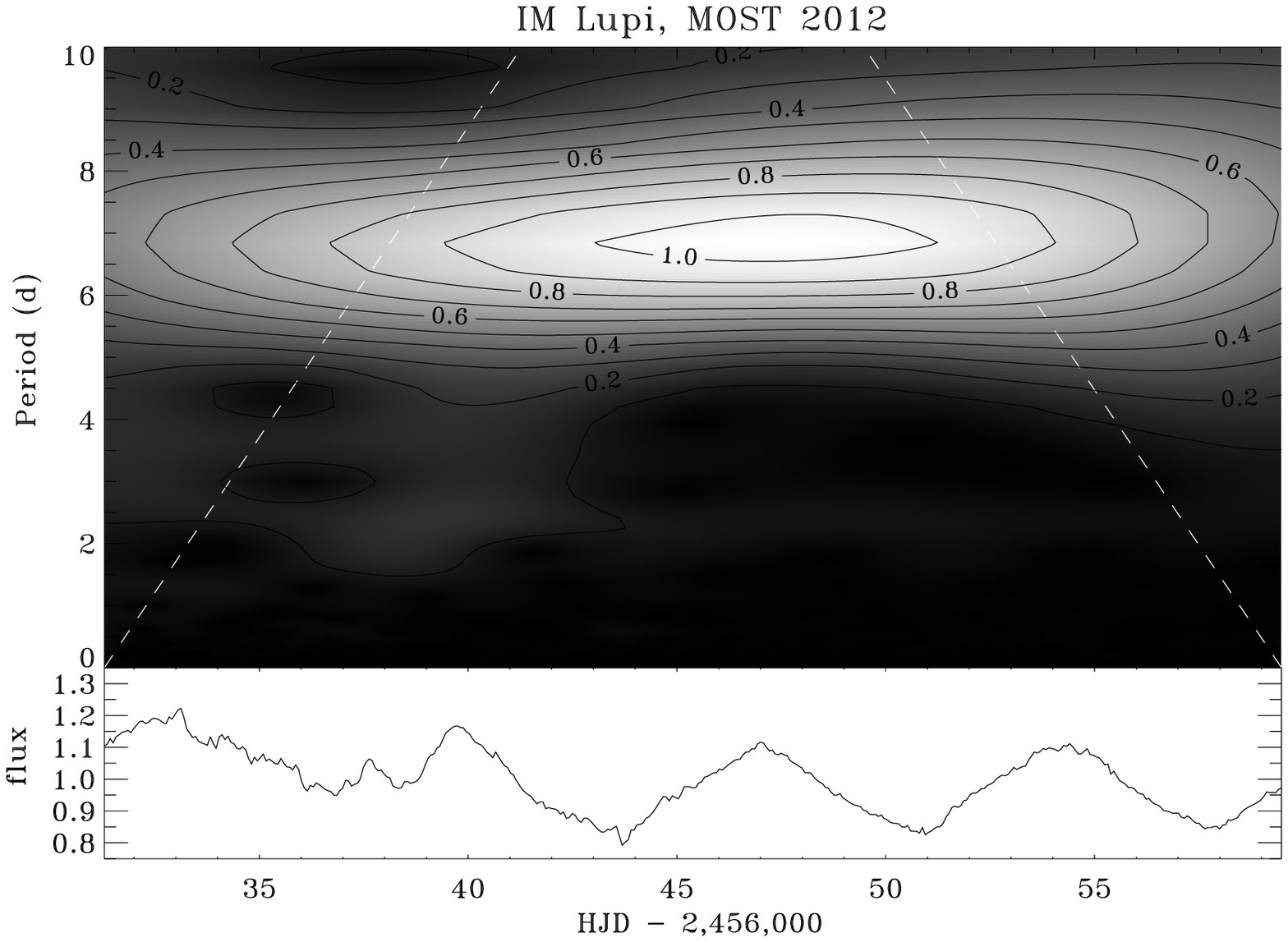}
\includegraphics[width=85mm]{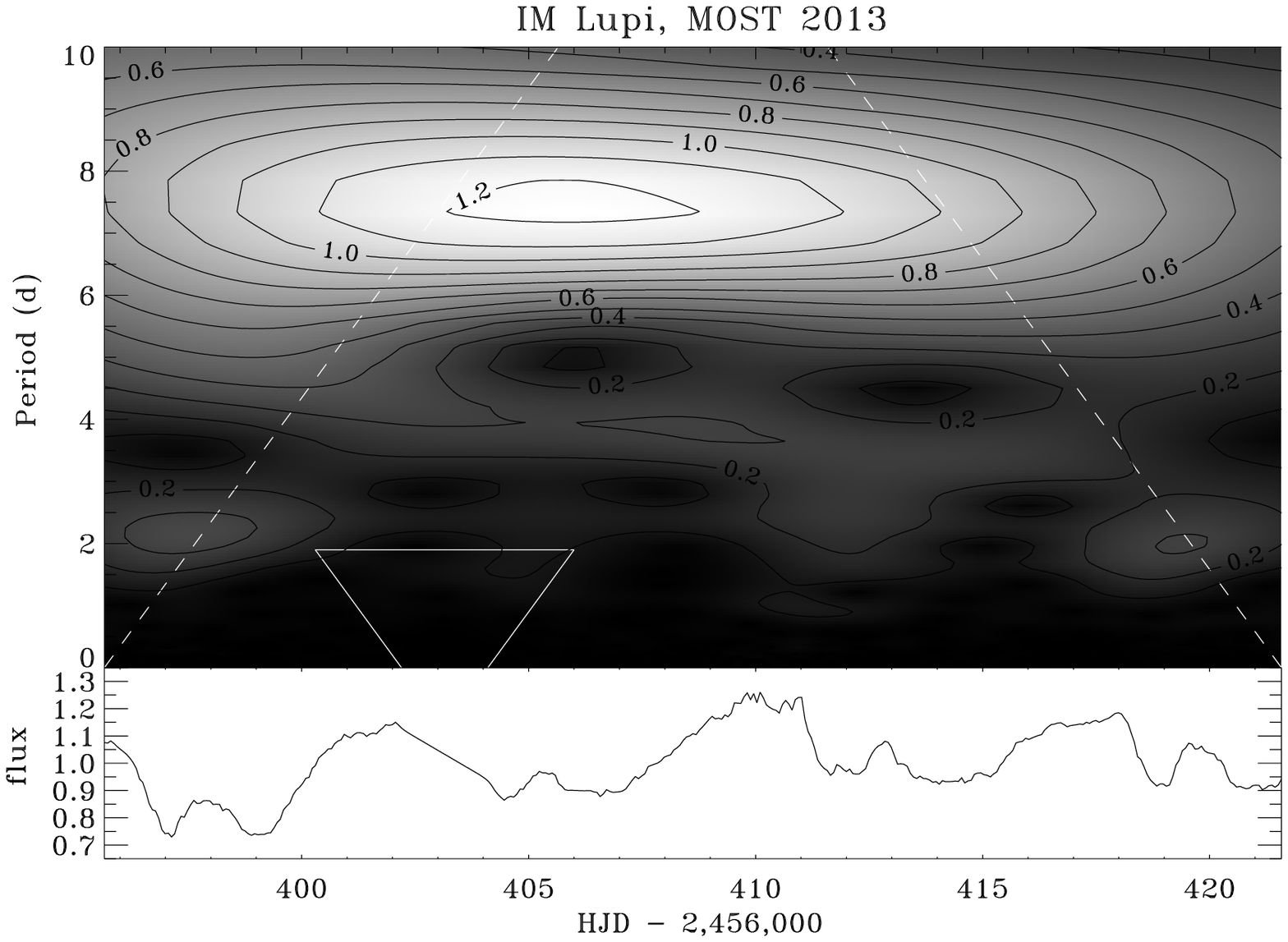}
\caption{The Morlet-6 wavelet transforms of the 2012 and 2013 
light curves of RU~Lup (upper panels) and IM~Lup (bottom panels). 
The amplitudes of the transforms are expressed by grey scale 
intensities and black contour lines. 
Primary edge effects are present outside the white broken lines 
but they do not affect our conclusions. 
Secondary edge effects, caused by 
the considerable 1.9~d break in the 2013 {\it MOST} data of RU~Lupi, 
are contained inside the small trapezoid defined by the three white solid 
lines and the X-axis. 
The bottom panels show the {\it MOST\/} light curves in normalized flux units, 
the same as in Figure~\ref{Fig.dat}.
}
\label{Fig.wav}
\end{figure*}

To obtain a uniform data sampling required for the    
wavelet analysis of RU~Lup and to remove a few interruptions 
in the data acquisition (see Sec.~\ref{mostobs}), 
we interpolated the 368 (2012) and 334 (2013) 
satellite-orbit, average data points into a grid of 401 and 
369 equally-spaced points at 0.07047~d intervals, for
the two seasonal {\it MOST\/} runs, respectively.
As in our previous investigations, the Morlet-6 wavelet 
provided the best match between the time-integrated power 
spectrum from the wavelet transform and the directly-obtained
frequency spectrum of the star, as given in the previous
section (Sec.~\ref{MOSTfr}).

The wavelet spectrum of 2012 light curve of RU~Lup (Fig.~\ref{Fig.wav}, 
the upper-left panel) shows at least 
two drifting quasi-periodic features, which are very
similar to those found in TW~Hya during 
the phases of unstable accretion \citep{ruc08, siwak11a}. 
The most pronounced is the feature which shortened its 
period from $\sim4.5$ to $\sim2$ d -- the feature can be also 
visually distinguished and tracked in the light curve from 
the beginning of the observations until $HJD\approx2,456,045$. 
The second feature is visible through the entire run and 
appears to shorten its period 
from $\sim5$ to $\sim4$~d by the end of the run.
We also note a few short-living $P=1-2$~d features, 
which appear and disappear on a time-scale of a few days.

Interestingly, the 2013 wavelet spectrum of RU~Lup does not show 
any clear period drifts for the quasi-periodic features.  
It is dominated by rather stable $P=1-2$~d features with
 life-time of a few days. 
Only the second half of the light curve contains 
a strong $\sim4$~d quasi-periodic signal.
We stress, however, that the 1.9~d long break in the data, which 
happened to take 
place during the most pronounced short-periodic oscillations
has a detrimental influence on the wavelet spectrum; thus 
the features visible inside and close to 
the small trapezium region marked by white continuous lines in  
Fig.~\ref{Fig.wav} represent numerical artifacts only. 
Thus, we cannot fully exclude the possibility that the period
drifts were present also in the 2013 data but were suppressed
by the presence of the data-taking gap. 
We also note that none of the longer time-scale features
(around $\approx8$~d) visible 
in both wavelet spectra can be regarded as a real effect 
due to relatively short duration of the observations; such
signals are not visible directly in the light curves.

We also analysed IM~Lup light curves in the same way as for RU~Lup.
As expected, the features representing the primary 
$\sim$7.5~d peaks in the Fourier spectra and directly visible in the light curves 
are also prominent and stable in the wavelet spectra 
(Fig.~\ref{Fig.wav}, bottom panels).

\subsection{Multicolour {\it SAAO\/} observations}
\label{saaoobs}

The {\it MOST\/} satellite is equipped with one broad-band filter.
Supplementary multicolour observations allow one to differentiate between effects 
caused by mechanisms suggested to explain T~Tauri star
variability: variable extinction, 
magnetic cold spots or hot spots produced by 
magnetospheric accretion. 
This is particularly important for IM~Lup, for which no 
previous multicolour light curves were available and for which
contradictory WTTS vs.\ CTTS classifications were proposed in the past 
(Sec.~\ref{imlup}). 
For this star a WTTS type is 
suggested by the 2012 {\it MOST\/} light curve, which looks similar to  
other WTTS targets observed by {\it MOST} \citep{siwak11b}. 
For this purpose we obtained multi-colour photometry 
of both stars at {\it SAAO} 
during 22 nights between 10 April and 7 May 2013. 
The stars were observed as secondary targets following TW~Hya, which was 
our primary target during the {\it SAAO} observations.

\begin{figure*}
\includegraphics[height=60mm,angle=0]{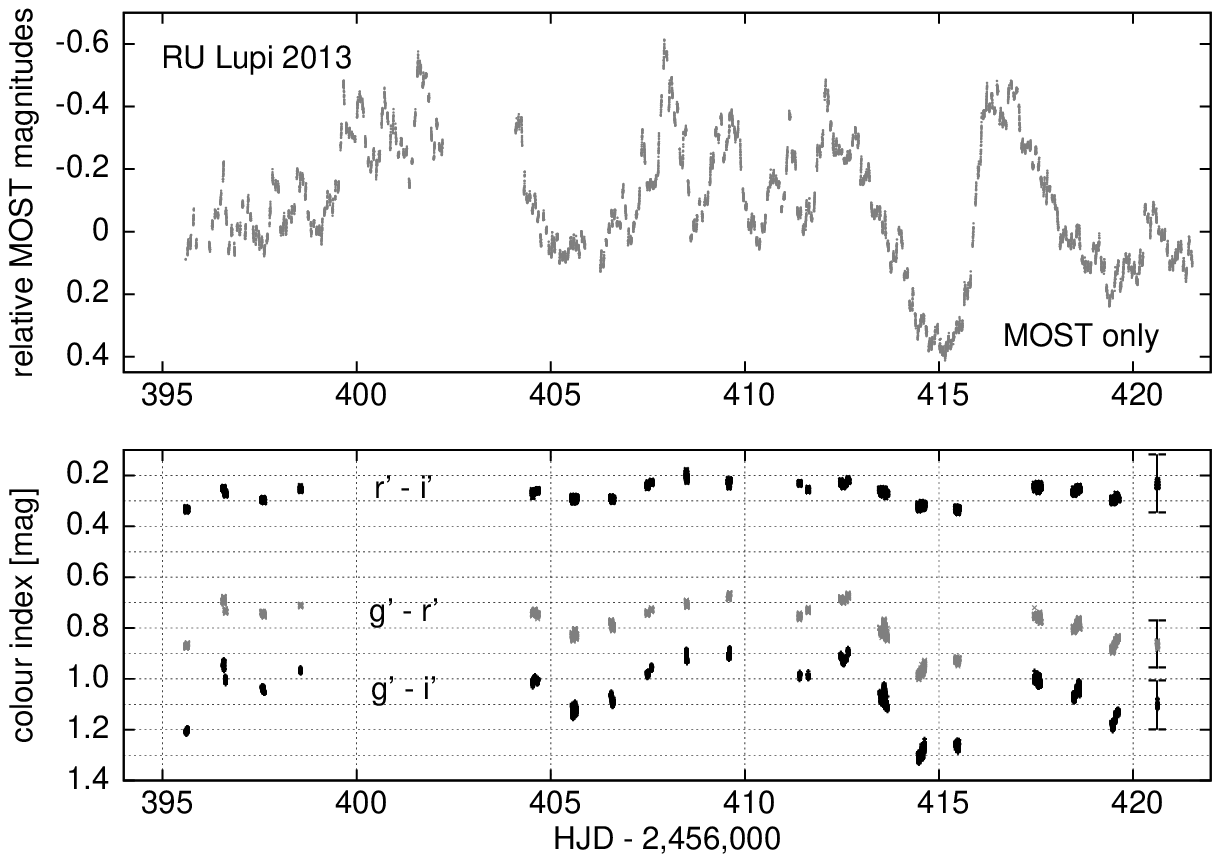}
\includegraphics[height=60mm,angle=0]{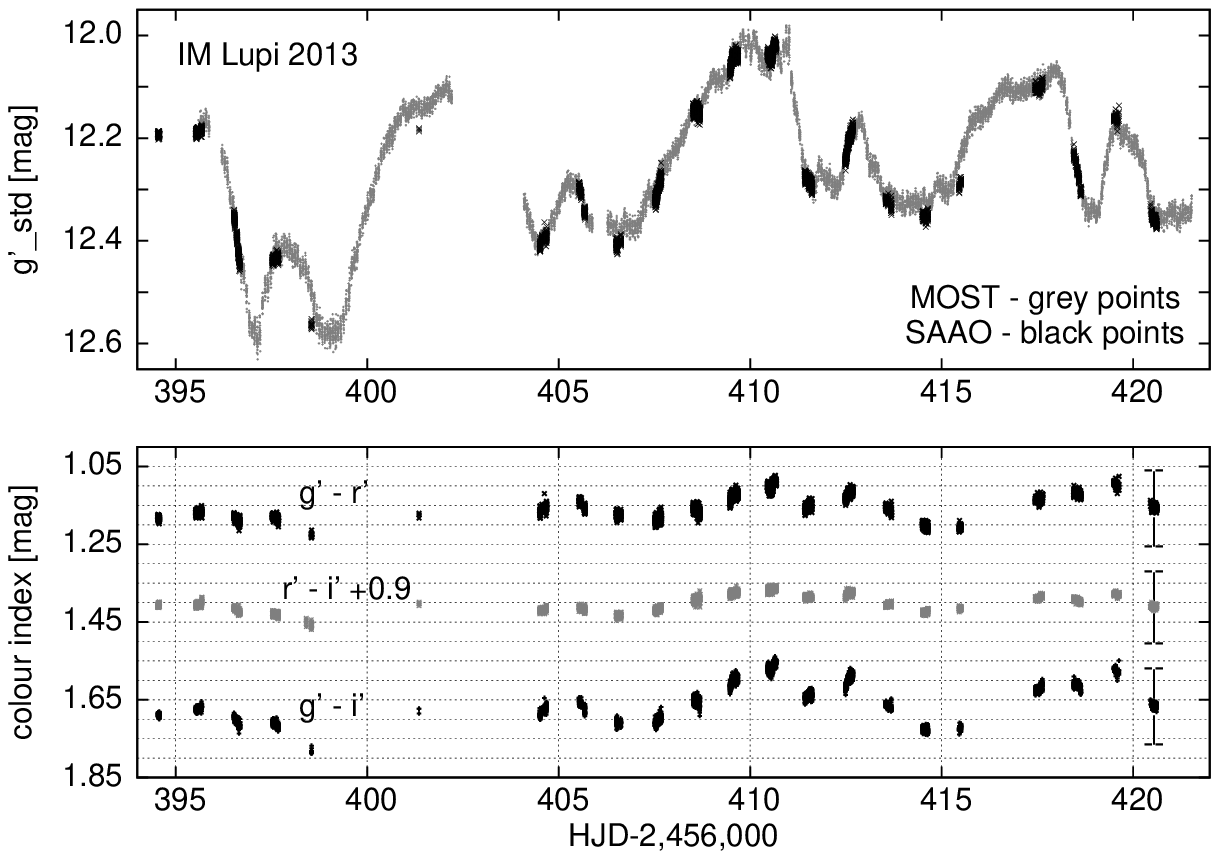}
\caption{
Colour-index variations of RU~Lup observed in 2013 at the SAAO are shown 
on the left panel. Since an independent light curve for RU~Lup
could not be extracted from the data, 
only the uncalibrated {\it MOST\/} light curve in 
magnitudes (grey points) is plotted in the upper panel 
for comparison of the light and the colour index variations. 
The right panel shows the Sloan $g'$-band light curve of 
IM~Lup (dark points) superimposed on 
the {\it MOST\/} light curve (grey points). Note, however, 
that the {\it MOST\/} magnitudes are expressed in
an arbitrary system and have been globally adjusted here
to the $g'$-band data.
The bottom panel shows the colour index variations in IM~Lup; 
the $r'-i'$ index has been shifted here by 0.9~mag for clarity. 
Note that for both stars the error in absolute colour index levels 
amounts to 0.1~mag (see Sec.~\ref{saaoobs}) -- for clarity, the error bars 
are drawn for the last night of observations only.
} 
\label{Fig.mult}
\end{figure*}

We used the 0.75~m telescope equipped with {\it TRIPOL} 
which is a visitor instrument of the Nagoya University 
donated to {\it SAAO}. 
The observed field of view of $3 \times 3$ arcmin was 
split by two dichroic mirrors and then directed to 
three SBIG ST-9XEI cameras through the Sloan $g'r'i'$ filters, 
resulting in three simultaneous images for each exposure. 
In addition to three times higher observing cadence, 
the crucial advantage of this instrument is the possibility 
to track colour-index variations even during non-photometric 
conditions which prevailed during our observations.
The typical spacing of our observations was 18--23~sec 
(including 3~sec read-out time), 
with full range of 13--33~sec, depending on the actual 
seeing and sky transparency. 
All frames were {\it dark} and {\it flat-field} calibrated 
in the standard way within the {\small \sc ESO-MIDAS} package \citep{Warmels91}. 
For photometric reduction we used the {\small \sc C-Munipack} 
programme \citep{Motl11} utilising the {\small \sc DAOPHOT~II} package.

We were not able to monitor brightness variations of RU~Lup
due to many non-photometric nights during our run. 
This star is isolated on the sky (about 15~arcmin from IM~Lup), 
without any suitable comparison star within 
the instrumental small field of view. 
Thus, only high-quality measurements of the 
colour index were feasible (Fig.~\ref{Fig.mult}, the left panel).
In the case of IM~Lup, nearby GCS~07838-00926 
served as a convenient comparison star; it was found to be 
constant in the {\it MOST\/} observations 
(Sect.~\ref{mostobs}). 
We derived its approximate Sloan magnitudes
on the night of 22/23 April 2013 using standard-star 
observations of the E600 region \citep{menzies89}
and the Johnson-system magnitudes of the standard stars transformed 
to the Sloan system using the equations given by \citet{fukugita96}:
$g'_{std}=12.54(2)$, $r'_{std}=11.34(4)$ and $i'_{std}=10.88(4)$. 
The light curve of IM~Lup in the $g'$-band obtained by referencing to this star
is shown in Figure~\ref{Fig.mult}. 
During the same night of photometric quality we frequently alternated observations of RU~Lup and IM~Lup, 
as used in the past in the era of one-channel photometers. 
This way we obtained an absolute calibration of RU~Lup's colour indices 
using the interpolated observations of GCS~07838-00926 as reference points.

A comparison of the RU~Lup colour indices to that of
a non-reddened Main-Sequence M0V star 
($g'-r'\approx1.35$, $r'-i'\approx0.60$; \citealt{fukugita96}), 
shows that the star is permanently bluer.
The same situation is observed for IM~Lup.
At first glance the blue-colour excess 
is well visible during the maxima and branches. 
During light minima the colour index are $g'-r'\approx1.23$ 
and $r'-i'\approx0.55$, and thus are bluer by only $0.13\pm0.09$~mag 
and $0.05\pm0.10$~mag than in M0V star. 
Note, that the errors are large but they include all uncertainties related 
with colour transformation to the Sloan system. 
However, the blue colour excess in IM~Lup was never observed 
by us to decrease to zero, 
which may be explained by geometrical circumstances,
such that before 
the hot spot located on the hemisphere directed towards the observer 
sets completely, the antipodal spot rises at the same time. 
We note that the colour indexes are significantly bluer  
in RU~Lup than in IM~Lup, which can be explained by the difference in the
mass accretion rate of some three orders of magnitude between the two
stars. 

The results are in accordance with the historical 
results (Sec.~\ref{rulup}) confirming that accretion and 
hot spots play major roles in RU~Lup's light variations.
The same conclusion can be formulated for IM~Lup for the first time 
based on results obtained in this work.

\subsection{ASAS data}
\label{asasobs}

\begin{figure*}
\includegraphics[height=60mm,angle=0]{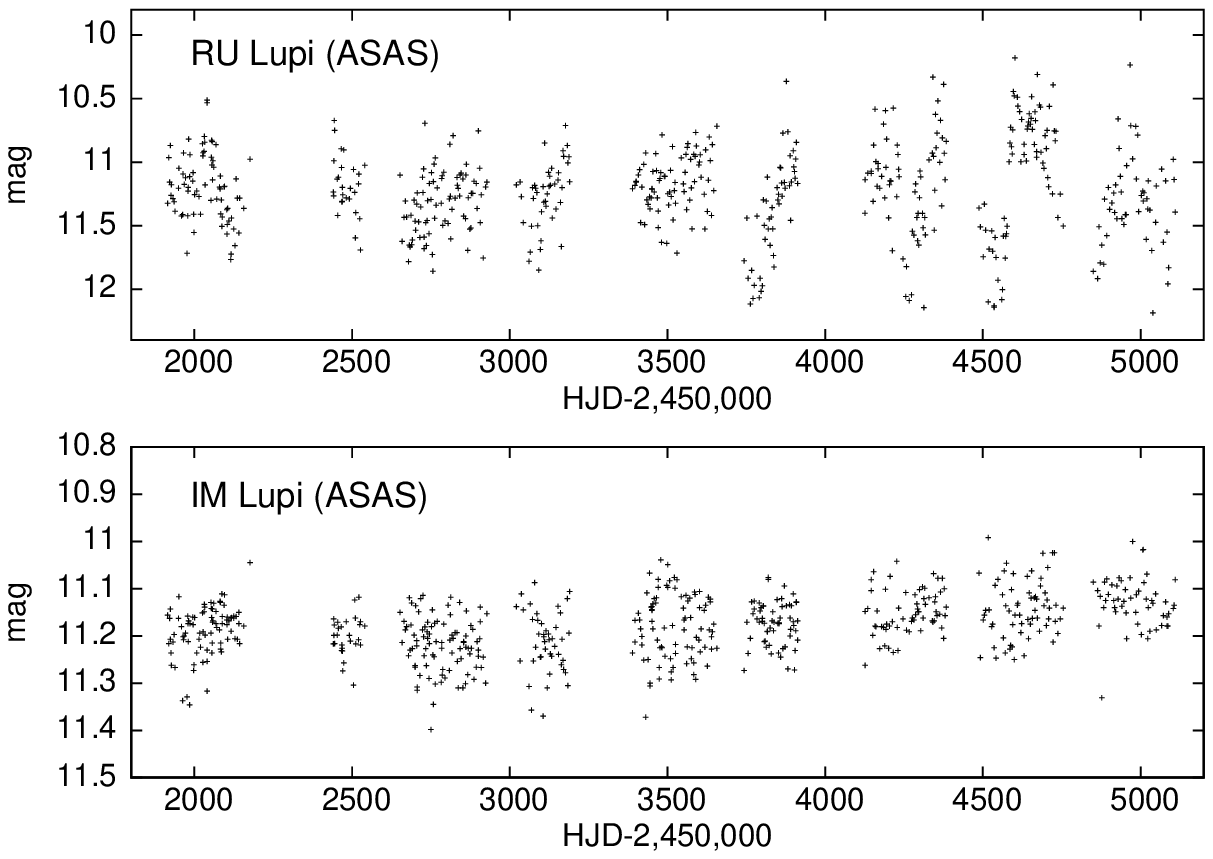}
\includegraphics[height=60mm,angle=0]{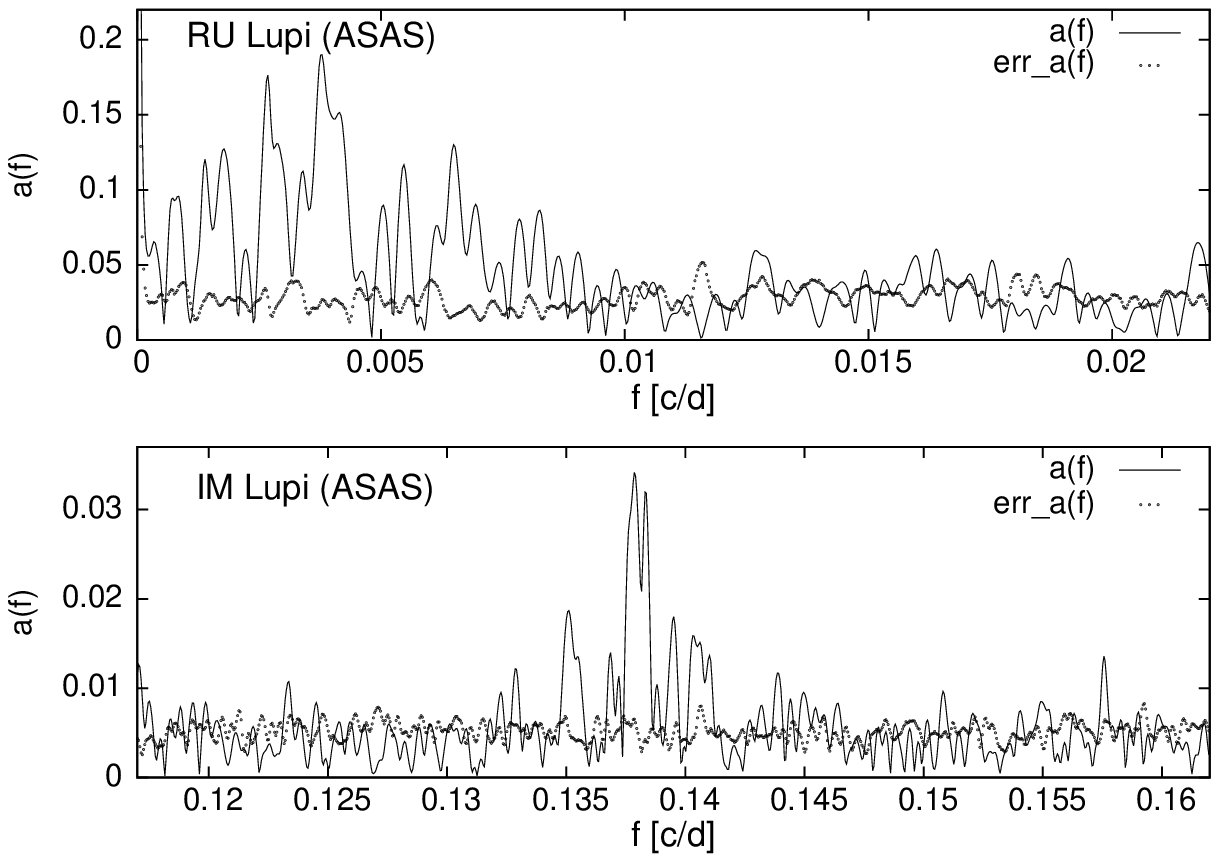}
\caption{The left panels show the long-term light variations of RU~Lup and IM~Lup 
observed by {\it ASAS} in the 2001-2009 seasons, while the right panels show the corresponding 
amplitude frequency spectra $a(f)$. The amplitude errors are marked by black dots.
}
\label{Fig.datasas}
\end{figure*}

While the new {\it MOST\/} and {\it SAAO\/} observations addressed the
variability of RU~Lup and IM~Lup in time scales of hours to about a week,
an analysis of the archival {\it ASAS} data
 provided supplementary information on
variability in long time-scales of days to years. 
The left-hand panels in Figure~\ref{Fig.datasas} show the 
{\it ASAS} Johnson $V$-filter light curves of 
RU~Lup and IM~Lup gathered during nine observing seasons 
from 2001 to 2009. The observations were obtained at intervals 
of typically one to seven days. 
For details of the data collection and reduction, see \citet{pojmanski04}.
The typical error of a single data point was 0.037~mag for both stars.

The average brightness level of RU~Lup was $V=11.225$~mag. 
The amplitude of light variations was moderate, 
$\sim0.35$~mag and constant until $HJD \approx 2,453,600$, i.e.\ the 
2005 season. Starting in 2006, the light variations 
were dominated by a long-term quasi-periodic oscillation 
with amplitude gradually growing with time up to $\simeq 0.9$~mag.
In contrast,
IM~Lup was fairly constant over the whole period of nine years, 
with the possibility of small sinewave-like variations. 
The average brightness level was $V=11.178$~mag. 
This value suggests that the visual companion GCS~07838-00926
(used as a comparison star for multi-colour observations, 
Sec.~\ref{saaoobs}) was included in the aperture during photometry extraction. 
The scatter in magnitude measurements, as measured by their 
standard deviation was 0.063~mag.

The Fourier spectrum of RU~Lup does not show any significant 
peaks at periods shorter than a month;
only the low frequency peaks grouped within 
$f=0.001-0.008$~c/d are significant
(Figure~\ref{Fig.datasas}, the upper-right-hand panel). 
In addition to the obvious signal at a period of 1~year 
($f=0.00274$~c/d), 
the strongest peak is produced by the well-defined 
quasi-periodic oscillation with $\sim0.9$~mag amplitude which 
has been present since 2006. 
The shape and the central frequency of the corresponding peak are 
thus better defined in the Fourier spectrum for the 
{\it ASAS} data after $HJD\approx2,453,600$. 
The period of the oscillation is $258\pm3$~d. 

In contrast to RU~Lup, the Fourier spectrum of IM~Lup shows a well 
defined strong peak at frequency 
$f=0.1380\pm0.0005$~c/d ($P=7.247\pm0.026$~d).
This value is very similar to the values of 7.19-7.58~d 
obtained from the {\it MOST} data (Section~\ref{MOSTfr}).

\section{High-resolution spectroscopic observations}
\label{spectra}

We analysed public-domain spectra of RU~Lup and IM~Lup 
in an attempt to broaden the scope of our photometric 
investigation based on the {\it MOST\/}, {\it SAAO}, 
and {\it ASAS}  observations.
For the spectral analysis of the velocity field,
we used the Broadening Function (BF) method \citep{ruc02}.  
The goals were: (1)~a comparison with the results utilising the 
cross-correlation function (CCF) analysis which 
reported possible spectral-line asymmetries in RU~Lup,
caused by photospheric spots \citep{stempels02, stempels07}
and (2)~a new study of the same line-asymmetry in IM~Lup.

The BF technique was introduced for the study 
of close binary stars some time ago \citep{ruc92} and 
since then has been considerably improved.
In many details, the BF method is very similar 
to the least-squares deconvolution ({\it LSD}) method
introduced by \citet{DC1997,Donati1997}.  
The BF approach has several advantages over 
the CCF \citep{ruc02}, mostly as a strictly linear method. 
It was extensively 
used during the David Dunlap Observatory radial velocity (RV) 
survey of short-period 
binary stars, as summarised by \citet{ruc2010,ruc2012}
proving to be an excellent method for studies of strong
line-broadening due to rotation, binarity or spottedness. 
It has been also applied for studies of matter 
flows between components of contact and near-contact binary stars 
\citep{pribulla08, siwak10, ruc15} as well as in an attempt at 
metallically determination from extremely broadened 
spectra \citep{ruc13}. 
In essence, the BF technique determines the Doppler
broadening kernel in the convolution equation transforming 
a sharp-line spectral template into the observed spectrum. 
This way any radial-velocity-field induced effects are isolated
as the kernel of the transformation.

\begin{table*}
\caption{Results for the Broadening Function analysis of 76 
archival {\it UVES} spectra of RU~Lup from years 2000 -- 2005 
and 2012, and 4 archival {\it HARPS} spectra of IM~Lup from 2008.
The spectral regions and velocity steps $\Delta v$,
used during Broadening Function determinations in two spectral 
regions of RU~Lup, and four regions for IM~Lup,  
are shown in the upper part of the table.  
$\sigma$ is the Gaussian smoothing used in the BF derivation  
to match the spectrograph's resolutions while $\Delta v_{\rm eff}$ is
the effective velocity resolution of the final BFs. 
Averaged values of radial velocities $\overline{v_{\rm rad}}$ and 
rotational velocities $\overline{v\,\sin i}$ 
obtained 
from Gaussian fits 
to RU~Lupi's BFs, and synthetic rotational profiles fitted 
to IM~Lupi's BFs are shown in the bottom lines.
}
\begin{tabular}{c c c c c c c }
\hline\hline
parameters/star & RU Lup 2000--2005 & RU Lup 2012 & IM Lup {\it blue}  & IM Lup {\it green}  & IM Lup {\it yellow} & IM Lup {\it red}  \\ \hline
spectral region [\AA]          & 5907--6041 & 5660--5948 & 4445--4772 & 4975--5300 & 5400--5728 & 5910--6208 \\
$\sigma$ [pix]        & 1.5        &  1.5       &     2.0    &   2.0      & 2.0        & 2.0        \\      
$\Delta v$ [km/s/pixel]       & 0.84       &  0.76      &    0.65    &  0.58      & 0.54       & 0.50       \\ 
$\Delta v_{\rm eff}$ [km/s] & 2.97       &  2.68      &    3.06    &  2.73      & 2.54       & 2.35   \\ 
\hline
Number of spectra       & 46    & 30     &   4      &    4              &   4           &   4               \\ 
\hline
$\overline{v_{rad}}$~[km/s]   & $-1.2(2)$  & $-1.2(2)$  & $-1.20$    & $-0.60$   & $-1.51$     & $-1.44$     \\
$\overline{v\,\sin i}$~[km/s] & 8.63(29)   & 7.33(8)    & 15.57      & 15.83     & 16.50        & 16.03      \\ 
\hline\hline 
\end{tabular}
\label{Tab.BF}
\end{table*}

\subsection{The spectral data and their analysis}

We made use of the Phase-3 {\it ESO\/} archival data
reduced by dedicated {\it ESO\/} pipelines or by actual observers.   
For RU~Lup we used 76 high-resolution spectra 
from years 2000 -- 2005 (46) and 2012 (30) obtained by the Ultraviolet Visual 
Echelle Spectrograph {\it UVES} \citep{Dekker04} during the programs identified 
as ID~65.I-0404, 69.C-0481, 075.C-0292 and 089.C-0299.
For IM~Lup, the data sets that we used consisted of four 
high-resolution spectra obtained with 
the High Accuracy Radial velocity Planet Searcher 
{\it HARPS} \citep{Mayor03} during two nights 
on 8/9 and 9/10 May, 2008 within the programme ID~081.C-0779.

Metallic absorption lines of RU~Lup below 
$\sim5660$~\AA ~are seriously contaminated 
by numerous and strong emission lines. 
Only the red spectral regions (5907--6041~\AA ~for the 2000--2005 data),
the same as previously used by \cite{stempels02} and \cite{stempels07},
are suitable for BF determination. 
Due to a slightly different setup of the spectrograph during 
the 2012 run, the BFs were determined from a
slightly bluer spectral region ($5660-5948$~\AA) 
than in the years 2000--2005 (Table~\ref{Tab.BF}). 
For all series of observations of RU~Lup, two fragments containing 
emission lines (within $5865-5910$ \AA\ and $5985.5-6000$ \AA) 
were omitted from the BF determinations. 
The condition of absolute absence of emission features in the process 
of BF determination is very important, as they can 
lead to false features in the BFs which are
similar to those produced by cold-spots
(genuine photospheric intensity depressions); 
we refer interested readers to Section~3 of \citet{ruc15} 
for more explanations. 

\begin{figure*}
\includegraphics[height=52mm,angle=0]{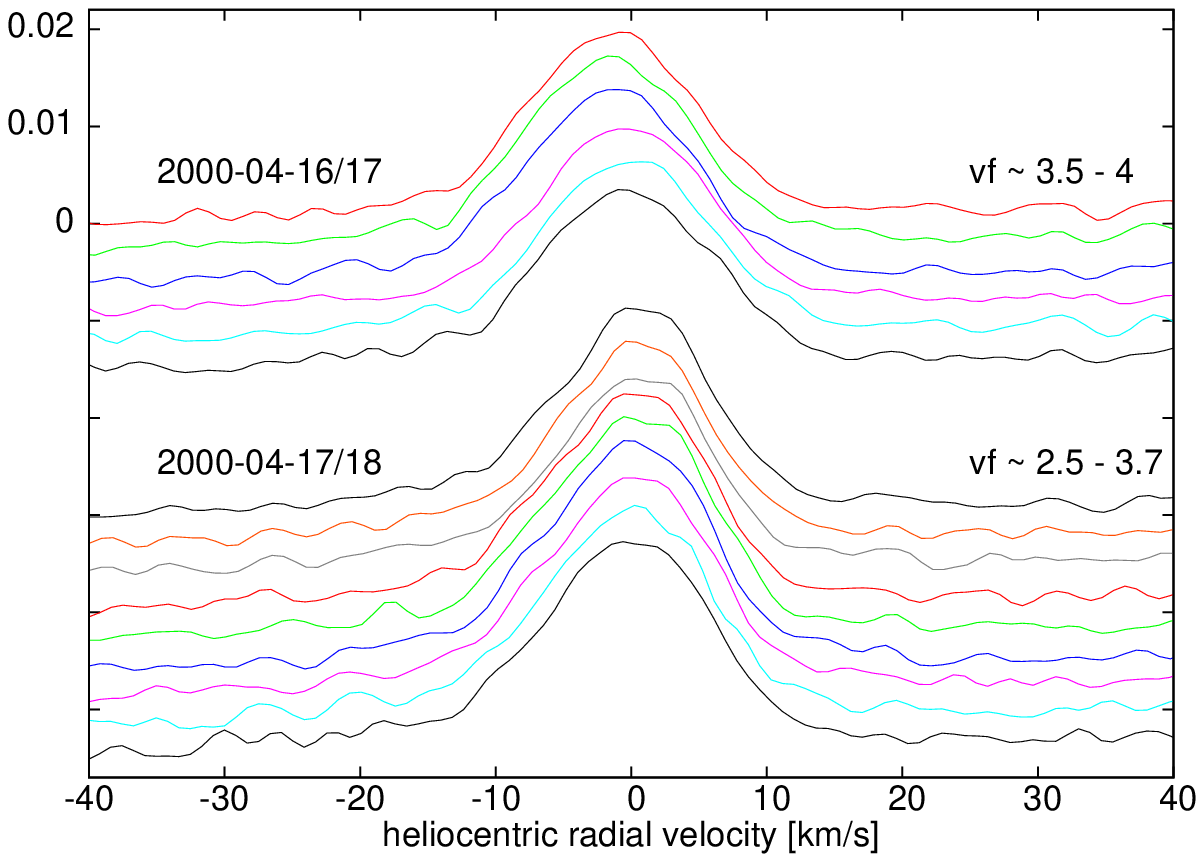}
\includegraphics[height=52mm,angle=0]{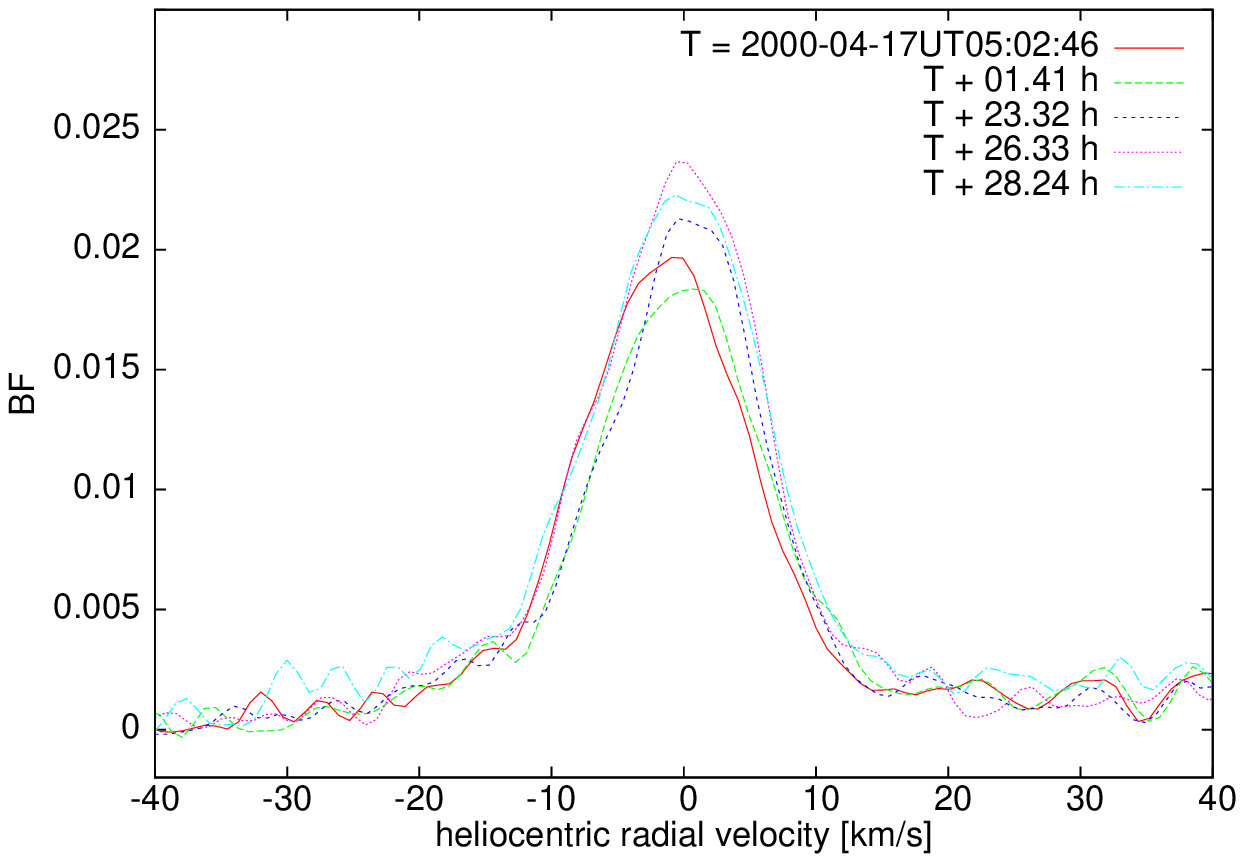} \\
\includegraphics[height=52mm,angle=0]{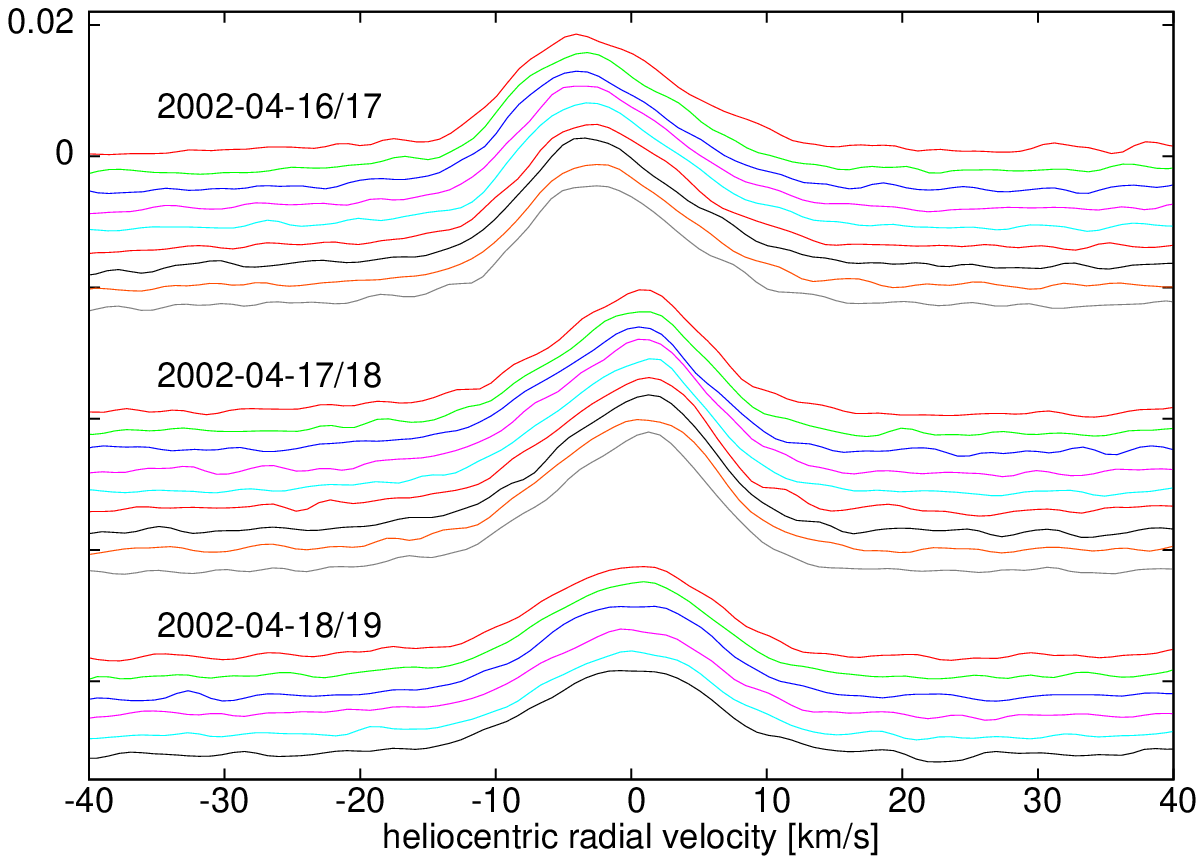}
\includegraphics[height=52mm,angle=0]{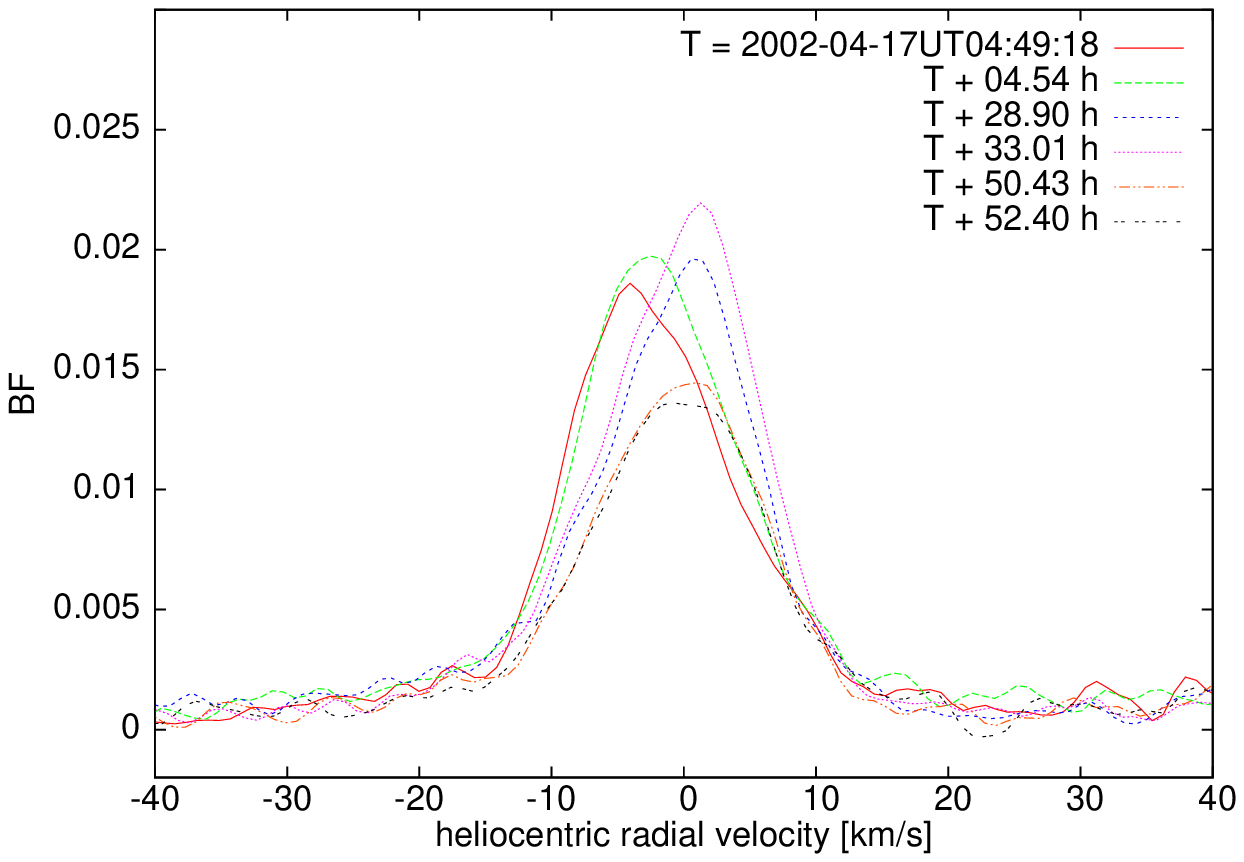} \\
\includegraphics[height=52mm,angle=0]{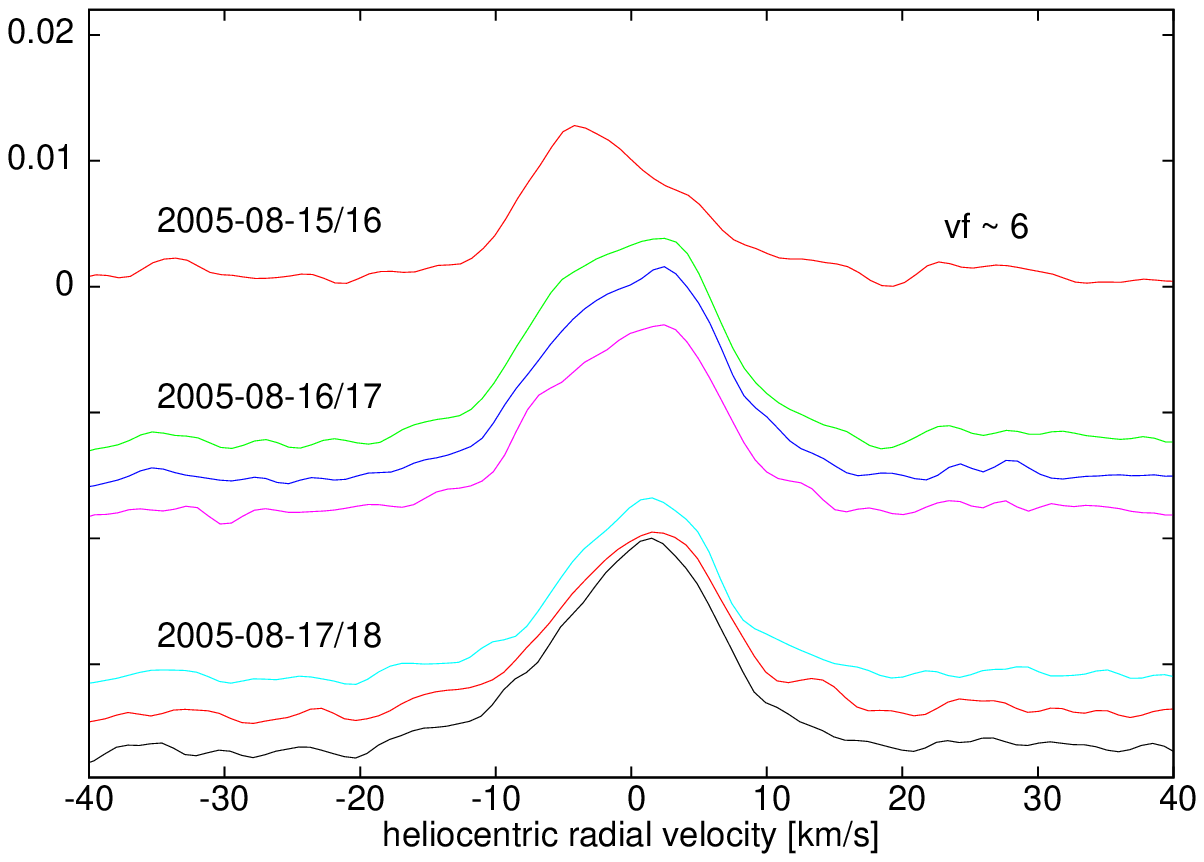}
\includegraphics[height=52mm,angle=0]{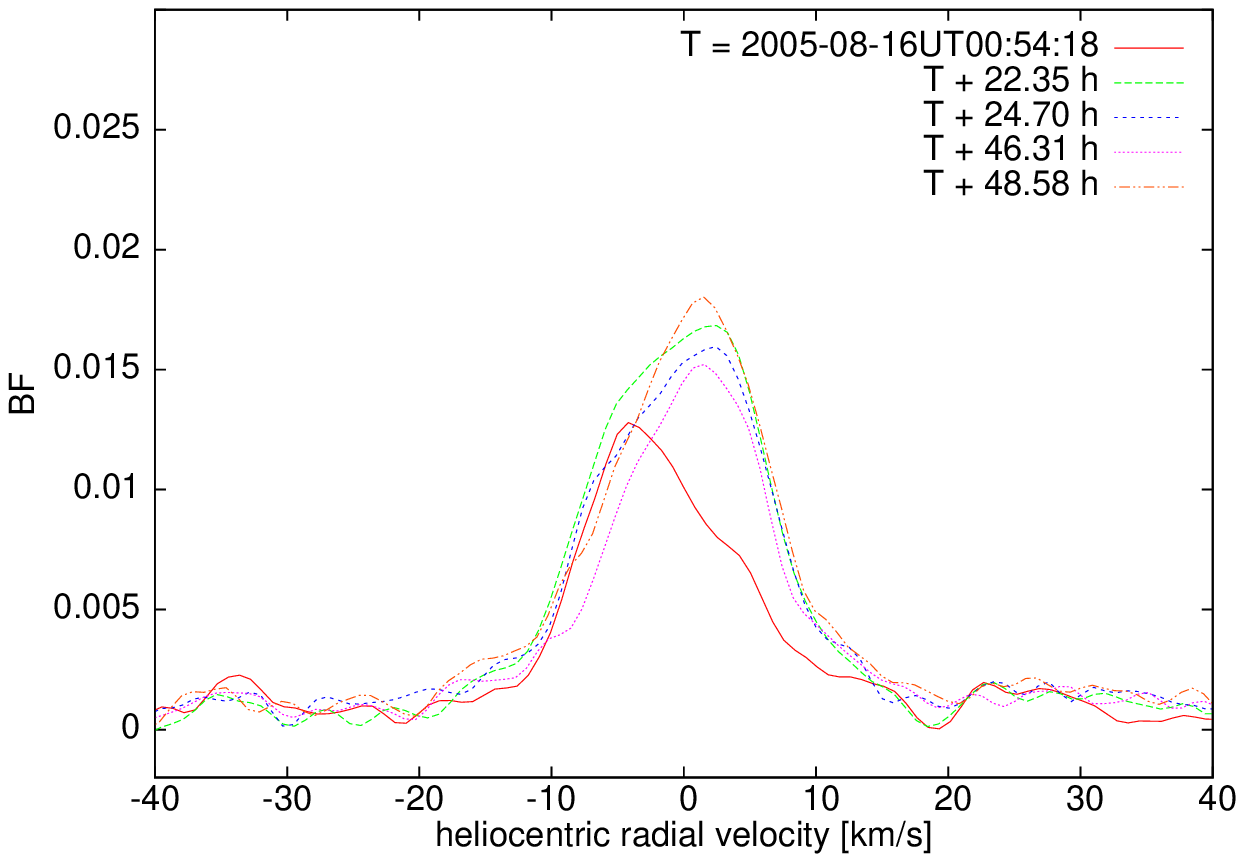} \\
\caption{The left panels show all 2000, 2002 and 2005 RU~Lup Broadening Functions 
and their temporal evolution. 
The first BFs obtained during each of the three seasons are plotted on top; 
observations obtained in consecutive nights are separated by vertical spaces. 
For some nights approximate veiling factors $vf$ are given. 
The right panels show some of the BFs presented in the left panels, selected
to emphasise 
variations in short time-scales of hours to days. 
The UT time T of the first BF is given to one second as the first
item in the legend while the time differences in hours from the first BF are shown 
for the next consecutive BFs. 
} 
\label{Fig.BF}
\end{figure*}

During the process of BF determination we used for both stars 
a single synthetic spectrum 
calculated with the programme {\sc SPECTRUM} 
\citep{gray10} using Kurucz's M0V model \citep{kurucz92} 
with solar abundances given by \citet{anders89}, 
with $\log g=4.0$ to match lower gravity of T~Tau stars.
The original BF's were smoothed with a Gaussian with a standard deviation  
$\sigma$ pixels to match the spectrograph resolution. 
This operation changed the original velocity sampling $\Delta v$ of the BFs 
(determined by the CCD pixel size) into the instrument-limited effective velocity resolution 
$\Delta v_{\rm eff}$, as given in Table~\ref{Tab.BF} and shown in Fig.~\ref{Fig.BF}, 
\ref{Fig.BFRU} and \ref{Fig.BFIM}.
The exact values of the smoothing parameters 
$\sigma$ turn out to have secondary significance 
on the details visible of the final BFs. 
We stress that the noise within individual BFs 
(measurable at the continuum level, outside the main peak) 
is significantly smaller than large-scale variations  
in the BFs reflecting surface heterogeneities, as
discussed in the next sections. The observational 
noise (from noise in the spectra as well as from
the BF determination) can be neglected at least to first approximation 
in the analysis.

Spectra of IM~Lup exhibit only moderate and weak Balmer and helium emission 
lines with absorption lines practically uncontaminated by 
emissions in their cores. 
This opens up the possibility of obtaining BF's from four different 
spectral regions to check for any wavelength-dependency and
the internal consistency of the velocity determinations. 
The four spectral regions are defined in Table~\ref{Tab.BF}.

\subsection{BF results for RU Lup}
\label{res-rulup-bf}

We show the BFs for RU~Lup in Figure~\ref{Fig.BF}.  
They are ordered in a temporal sequence spanning years
2000 -- 2005 on the left side of the figure and for selected,
densely-covered, nightly intervals on the right hand side.
We measured the BFs to extract essential radial velocity indicators
such as the mean velocities or the degree of broadening.
It is instructive to
compare the results obtained with the BF and CCF methods \citep{stempels02, stempels07}.
The mean radial velocities of RU~Lup were obtained by fitting 
a Gaussian function that in parts 
were judged to be symmetric, almost perfectly matched to the BFs profiles. 
The mean radial velocities exhibit a jitter of
$1.7\pm0.3$~km/s, i.e.\ smaller than 2.17~km/s found 
by \citep{stempels07} using the CCF method.
The discrepancy can be understood by different  
sensitivities of both methods to various parts of the
velocity-broadened line profile: 
the CCF method is sensitive to the absorption line centres,  
which -- because of the line asymmetries caused by the spots -- exhibit 
more extreme variations than the entire spectral lines to which the BF method is more sensitive.
The mean radial velocity of the star extended over all spectra, $\overline{v_{rad}}$, 
derived from the BF method is $-1.2\pm0.2$~km/s, 
i.e.\ slightly larger (by 0.3~km/s) 
 than that determined by the CCF. 
We also find that using the BF method the average rotational line broadening velocity $\overline{v\,\sin i}$ 
values obtained from the 2000-2005 
and 2012 data sets, i.e. 8.63 and 7.33~km/s, respectively (Table~\ref{Tab.BF}), 
are smaller in comparison with $\overline{v\,\sin i} = 9.1\pm0.9$~km/s
determined using the CCF.

\begin{figure}
\includegraphics[width=80mm,angle=0]{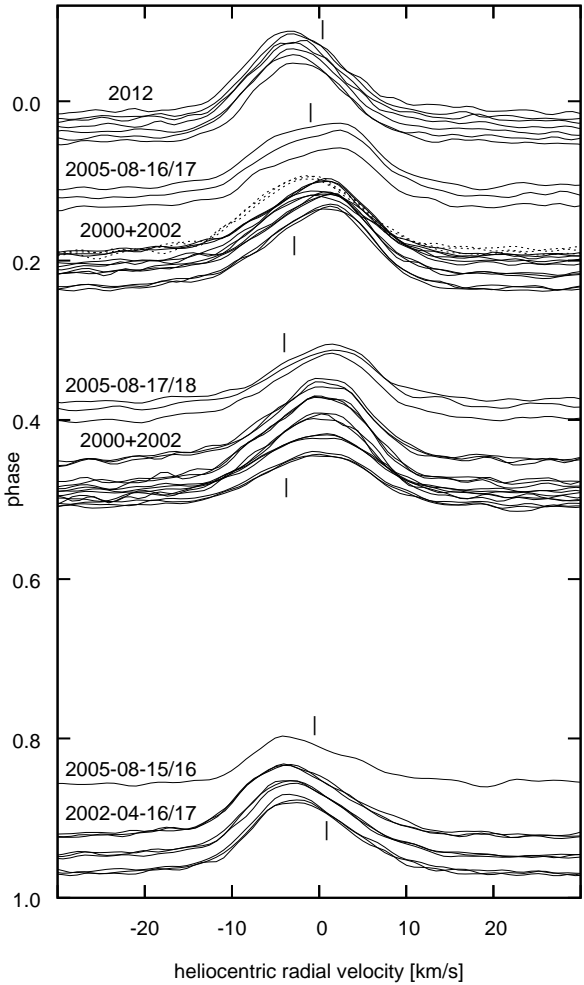}
\caption{Broadening Functions of RU~Lupi phased with period 3.71058~d, 
in arbitrary intensity units. 
The vertical marks indicate depressions in BF intensities 
explained by us as due to a large cold spot. 
The base of each BF was shifted from zero level by the
respective value of phase as on the vertical axis. 
Note that the first three BFs obtained during the
first night of 16/17 April 2000, show deviations from 
the overall tendency -- they are drawn by dotted lines for clarity (see also in Sec.~\ref{res-rulup-bf}).
}
\label{Fig.BFRU}
\end{figure}

The real advantage of the BF approach over the 
CCF for determination of the rotational period 
is visible in the left panels of Figure~\ref{Fig.BF}: 
in spite of the low value of $v\,\sin i$ the BFs directly show considerable asymmetries, 
i.e.\ intensity depressions that may be caused by 
the long-persisting, large polar cold spot postulated by \citet{stempels07}. 
The overall shape of BF profiles for RU~Lup appears to vary 
in a time-scale of about 4~d. This is particularly well visible during 
the three consecutive nights of April, 2002, as shown in Figure~\ref{Fig.BF}.
This leads us to confirmation of the result of \citet{stempels07}
since the radial velocities derived for
the whole time span of 13 years (including the 2012 data) 
can be perfectly well phased with the single period of 3.71058~d. 
We show this result in Figure~\ref{Fig.BFRU}, where all the BFs obtained over 
13 years are ordered by the rotational phase calculated for this value of the period. 
The initial epoch was set arbitrarily at 0~UT of 16 Apr, 2000, i.e.\ the day 
of the first {\it VLT-UT2 UVES} RU~Lup spectrum. 
The depressions caused by the cold spot are marked with vertical dashes. 
Note that, given the pole-on visibility of RU~Lup ($i=24$~deg), 
the spot remains visible during the entire stellar rotation. 
Similar figures obtained with phases calculated with slightly 
different values of the period look absolutely chaotic with
none of the coherent variations in the BFs as clearly
defined as in Figure~\ref{Fig.BFRU}. 
Only the first three BFs obtained during the
first night of 16/17 April 2000 show deviations from 
the overall tendency -- they are drawn by dotted lines in this figure for clarity. 
We claim that either a small reorganization of the cold spot, or appearance 
of short-term hot spot(s) with effective temperature(s) of 4500-7500~K 
may produce extra intensity in these BFs at negative radial velocities 
(see also in Section~\ref{res-imlup-bf}). 

In addition to the asymmetries caused by the
changing visibility of a large cold spot,
we observe rapid -- lasting only a few hours -- variations 
of BF intensities due to variable spectral veiling which fills all spectral
lines with an additional component, reducing their depth.
Chaotic accretion bursts directly observable in the RU~Lup 
light curves (Fig.~\ref{Fig.dat}) occur on similar time-scales
and seem to be the main reason for these variations. 
The first panel of Figure~\ref{Fig.BF} 
shows two values of $vf$ interpolated 
from the values given in Fig.~3 of \cite{stempels02}, 
estimated from adjacent spectral regions to those used for the BF calculation. 
A very high $vf\approx6$ was also found by \cite{gahm13} 
in the first 2005 spectrum presented in the bottom-left 
panel of Figure~\ref{Fig.BF}.
As expected, due to the linear properties of BF, these values correlate with 
BF intensities in the sense that 
{\it higher veiling leads to smaller integrated BF intensity}. 
This opens up possibilities to re-determine 
veiling factors with BF intensity integrals; we will return 
to this issue in a separate publication, as it is beyond 
the scope of this paper.

\subsection{BF results for IM Lup} 
\label{res-imlup-bf}

\begin{figure}
\includegraphics[height=60mm,angle=0]{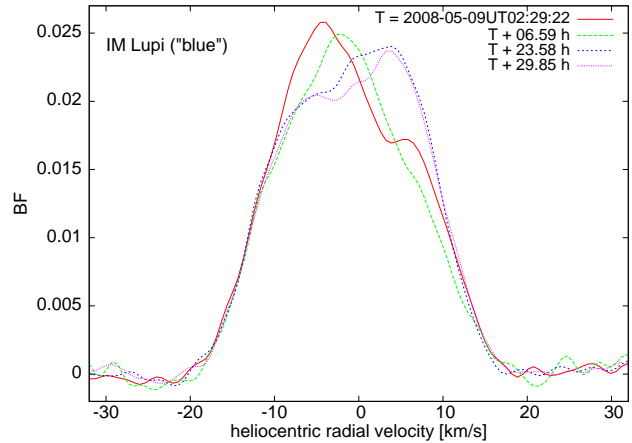}
\caption{Broadening Functions of IM~Lup obtained from 
four {\it HARPS} spectra in the {\it blue} 
spectral region (see Tab.~\ref{Tab.BF}). 
The BFs evolve as time progresses, 
which is mainly due to stellar rotation with most likely 
both hot and cold spots on the surface. 
} 
\label{Fig.BFIM}
\end{figure}

Temporal variations of the IM~Lup
Broadening Functions are shown in Figure~\ref{Fig.BFIM}.
The BFs were obtained in the {\it blue} region within 30~hours;
similar BFs obtained in other spectral bands change 
in a very similar way. 
Visual tracking of variations in 
consecutive BFs (see the legend of the figure to restore the time-line) 
suggests that the asymmetries are time-coherent and may be due 
to changing visibility of cold spots on the stellar surface. 
However, in contrast to RU~Lup, the reasons for detailed variations 
in the BF profiles are probably more complex:
IM~Lup does not show any obvious veiling (see Sec.~\ref{imlup})
and in many respects appears to be similar to MN~Lup, for
which hot spots of 5000--5800~K 
were detected on Doppler maps by \citet{strassmeier05}. 
We claim that photometric variability of IM~Lup 
is due to a hot spot temperature of $\sim6300\pm1000$~K, as 
crudely estimated from the first part of 2013 {\it MOST} 
data ($HJD-2,456,000\approx394-402$) 
by means of the recent version of the Wilson-Devinney light curve synthesis 
code \citep{wilson79}\footnote{In this process, we considered a grid of circular spots with 5--10~deg radii 
localized at latitudes 17~deg below the magnetic poles of the star. 
We took into account the known inclination, effective temperature of the star 
and the angle $\theta$, as estimated in 3rd paragraph of Section~\ref{discussion}. 
Equal temperatures of both polar hot spots were assumed and were searched manually 
to get the observed amplitude of light variations.}. 
Such hot spots may affect rotationally-broadened stellar 
absorption lines in a fragmentary way with extra absorption. This would affect the line shapes
appearing as a local intensity peak in the BF, which is otherwise
fully determined by the colder surface areas. 

We measured radial velocities of IM~Lup by 
fitting synthetic rotational profiles to the entire individual BFs.
Theoretical profiles were calculated using linear 
limb-darkening coefficients from tables 
of \cite{diaz-cordoves95} and \cite{claret95}, applied
separately for each spectral region.
Using $\overline{v\,\sin i}=15.98$~km/s, the value of the inner-disc 
inclination of $50\pm5$~deg, and assuming that it is co-planar with 
the stellar equator, we obtain practically the same value of the stellar 
radius $R=2.99_{-0.22}^{+0.40}$~R$_{\odot}$ as \citet{pinte08}.

\section{Discussion and Summary} 
\label{discussion}

Although the stars IM~Lupi and RU~Lupi
have been well studied from the ground (see Section~\ref{history}), 
the results obtained in this work
shed new light on details of accretion properties
in these CTTSs. 
Both stars fit well into the picture inferred from  
theoretical numerical 
simulations of magnetised-plasma 
accretion from the innermost accretion discs of 
\citet{romanowa04, romanowa08, kulkarni08, romanowa09, 
kulkarni09}.

The two stars analyzed in this paper display accretion phenomena
taking place in two different regimes 
of stable and unstable accretion.
IM~Lup, having a low mass-accretion rate (Sec.~\ref{imlup}) and 
fairly regular light curves modulated by hot spots with 
a period of 7.19 -- 7.58~d, is an excellent example of 
a CTTS accreting in the stable regime. 
In this regime, the steady plasma flow takes place 
from the inner disc toward the stellar magnetic poles 
in two funnels encircling the magnetosphere, to produce 
two antipodal arc-like hot spots which co-rotate with the star. 
Depending on the inclination angle and the misalignment 
angle $\theta$ between the stellar rotation axis 
and the magnetic pole axis, either one or both hot spots 
can be visible to an observer during a single stellar rotation. 
The lack of optical veiling in IM~Lup is in line with the moderate 
hot spot temperature of $6300\pm1000$~K, crudely estimated from the 2013 {\it MOST} 
light curve.

\cite{kulkarni13} considered shapes, widths and positions of hot 
spots created by stable accretion funnels. 
They found that for a broad range of parameters, the arc-like 
hot spots are formed at stellar latitudes 16--18~deg 
below the magnetic pole, and their typical polar widths are 7--8~deg.  
In combination with the finding from the {\it MOST\/} observations of IM~Lup, 
which in 2012 showed a fairly regular light curve likely modulated by a single spot, 
while in 2013 by two hot spots, the theoretical results can be used to estimate $\theta$. 
Assuming that the stellar equator is coplanar with the inner 
accretion disc, for rotation angle inclination of 
$i\approx50$~deg \citep{pinte08}, the spot on the opposite 
hemisphere may occasionally 
be invisible to the observer if $\theta < \approx 19$~deg, 
with some likely variations in the hot spot latitudes, their
polar widths or maybe even of $\theta$ itself.

The variability phenomena of RU~Lup are much more complex than in IM~Lup.
Due to the chaotic character of light variations and a lack of any single 
stable periodicity in light curves, 
we infer that the unstable accretion regime operates in RU~Lup.
As noted in Section~\ref{intro}, in this regime, 
which is favoured in CTTS having high mass accretion rates 
(see Sec.~\ref{rulup}), the hot spots are formed by a 
few tongues created by RT instabilities. 
The number, position and shape of the hot spots change 
on the inner disc dynamical timescale, sometimes as low as 1--2~d, 
masking the stellar rotation frequency in 
the frequency spectra \citep{romanowa08, kulkarni08, kulkarni09}. 
Hence, in spite of about 52 days of continuous photometric 
monitoring of the star with {\it MOST\/}, we were unable to detect 
any stable periodicity corresponding to the persisting 3.71~d signal 
in radial velocities discovered by \citet{stempels07}. This is
fully and independently confirmed by 
our re-analysis of archival spectra using the Broadening Function 
approach (Sec.~\ref{res-rulup-bf}). 
This 3.71~d signal most likely represents the true rotational period of the star.

During the process of unstable accretion, the 3D MHD numerical models predict 
the appearance of quasi-periods distributed around the value of the stellar rotational 
period. 
Such 4.5--2~d and 5--4~d quasi-periodic features were indeed noticed in the 2012 
light curve of RU~Lup (Sec.~\ref{wavmost}). 
The most extreme case, however, is represented by several 2--1~d quasi-periods detected 
in 2012 and 2013, which are apparently caused by hot spots moving relative to the photosphere 
with a considerably higher inner disc rotational frequency. 
In accord with \citet{kulkarni08, romanowa09, kulkarni09} the tongues formed through 
the RT instabilities originate at the inner disc radii of about 5.8~R$_\odot$, a value which 
is significantly smaller than the disc co-rotation radius of 8.74~R$_\odot$, evaluated from 
the stellar rotational period of 3.71~d and the mass of 0.65~M$_\odot$ (Sec.~\ref{rulup}).
We also note that the 2012 {\it MOST\/} observations of 
RU~Lup showed drifting quasi-periodic features, similar to TW~Hya 
observed in 2008 and 2009 \citep{ruc08,siwak11a}. 
This can be interpreted in the context of the results presented in 
Section~4.3 of \citet{kulkarni09}, which indicate that quasi-periodic
features may show a tendency towards period shortening for the 
increasing mass accretion rate within individual tongues. 
This leads to the non-linear development of RT instabilities 
and a local reduction of the inner-disc radius, what in turn produces a progressive 
increase of the rotational frequency of the tongue, and of the resulting hot spot. 
Although the quasi-periodic features observed in RU~Lup in 
2013 did not show any obvious trends in the period length, 
the apparent absence may be due to our inability to detect them 
through wavelets due to the long break in the data acquisition 
(see Sec.~\ref{mostobs} and in Fig.~\ref{Fig.wav}).\newline 

Whereas the day time-scale variability appears to have direct explanation 
in the accretion phenomena between the inner accretion disc
and the surface of the star, it is unclear what mechanism produces 
the long-term $258\pm3$~d quasi-periodic variation in RU~Lup. 
We saw this over the span of four years in the
{\it ASAS} observations (Sec.~\ref{asasobs}). 
The long time-scale suggests that the outer parts of the accretion disc where
some density disturbances occur, may indirectly lead to semi-periodic light variations. 
One possibility is pressure scale-height perturbations or proto-planetary nuclei 
that can create spiral arms \citep{juhasz15}.
We suggest that perhaps the spiral arms could lead to 
cyclical (i.e. the $\sim~258$~d) modulation of the inner disc plasma 
density and thus of the mass accretion rate, the hot-spot filling factor 
and associated veiling, which ultimately would affect the mean brightness 
of the star when averaged over monthly intervals. 
Unfortunately the {\it UVES} spectra were obtained during the 
first half of the {\it ASAS} observations, 
when the 258~d quasi-periodic oscillation did not seem to be present. 
Therefore we could not find any possible correlation 
between the mass accretion rate estimated from the spectra, and the mean 
stellar brightness from the available photometric data.

The more complex of the two stars, RU~Lup, is similar to TW~Hya, 
which is the best photometrically explored CTTS so far from space. 
Both stars have similar spectral type, show vigorous accretion, 
are visible in a pole-on orientation, and 
long-lasting cold spots near the magnetic poles modulate
their radial velocity variations with similar
periods of 3.71~d (RU~Lup) and 3.57~d (TW~Hya, \citealt{donati11}). 
The four {\it MOST\/} runs of TW~Hya provided light 
curves showing evidence of a constantly varying 
accretion regime \citep{siwak14}, in accordance with 
the \cite{romanowa08} prediction of alternating episodes of stable 
and unstable accretion, depending on the mass accretion rate. 
It is therefore not excluded that at times, perhaps during 
the brightness minimum, when $V$ drops down to $\approx12.8$~mag 
(see Section~\ref{rulup} and the previous paragraph), 
RU~Lup may show accretion running in a moderately stable or even a stable regime. 
Such events would allow one to determine the rotational period 
of the star from photometry \citep{kurosawa13}.\newline
Ideally, future space-based photometric observations should be followed 
by spectro-polarimetric observations capable of reproducing the hot spot 
distribution in stellar coordinates. 
For instance this technique applied to TW Hya in 2008, revealed polar, 
mid- and low-latitude brightness excess regions \citep{donati11}, indicating 
the presence of hot spots created both through a stable polar funnel and unstable tongues. 
In spite of this evidence of moderately stable accretion, the light curve simultaneously 
obtained by {\it MOST} \citep{ruc08} is dominated by clear signs of unstable accretion.

While long-term, continuous photometric monitoring of CTTSs from
space has now become a well established tool \citep{cody14,stauffer14}, 
such observations frequently show very complex superposition of many phenomena, particularly in T~Tau
stars accreting in the unstable regime, such as RU~Lup or TW~Hya.
For the determination of rotational periods, a much more efficient approach
is through the use of high-resolution spectroscopy followed
by analysis using linear deconvolution methods, such as for TW~Hya
by \citet{donati11} or for RU~Lup analyzed in this paper. 
Unlike photometry, such methods are sensitive to cold-spot induced 
line-profile variations permitting rotational period 
determination even from a moderate number of spectra. 
The radial velocity variations induced by such spots are not negligible
and reach amplitudes that may exceed 2~km/s, a limitation which 
carries an important ramification for RV planetary searches around CTTSs and
may explain the lack of success for the extant searches 
\citep{bailey12,crockett12,lagrange13}, since the required 
accuracy necessary for discovery of Jupiter-mass planet is 
of the order of 100~m~s$^{-1}$. 
Linear deconvolution techniques, such as the Broadening Function technique, 
and the classical Doppler imaging technique, may permit 
removal of the star-spot perturbations \citep{donati15}, particularly for the ``cleaner''
CTTSs accreting in a stable regime.

\section*{Acknowledgments}
\label{thanks}
This study was based on:\newline
(1)~data from the {\it MOST\/} satellite, a Canadian Space Agency 
mission jointly operated by Dynacon Inc., the University of Toronto Institute 
of Aerospace Studies, and the University of British Columbia, with the assistance 
of the University of Vienna,\newline 
(2)~observations made at the South African Astronomical Observatory,\newline 
(3)~data obtained from the {\it ESO} Science Archive Facility under
request numbers 145841 and 145843, within the pogrammes ID~65.I-0404, 69.C-0481, 
075.C-0292, 081.C-0779 and 089.C-0299,\newline
(4)~data obtained from the All Sky Automated Survey ({\it ASAS}) telescope of 
the Warsaw University Astronomical Observatory.\newline
MS and WO are grateful to the Polish National Science Centre 
for the grant 2012/05/E/ST9/03915. 
The Natural Sciences and Engineering Research 
Council of Canada supports the research of DBG,
JMM, AFJM and SMR. 
Additional support for AFJM was provided by FRQNT (Qu{\'e}bec). 
CC was supported by the Canadian Space Agency. 
RK and WWW is supported by the Austrian Science Funds (P22691-N16).
MS and WO acknowledge Dr.\ Hannah Worters and the entire {\it SAAO} staff 
for their hospitality.
This paper also made use of NASA’s Astrophysics Data System (ADS) Bibliographic 
Services.\newline 
Special thanks also go to the anonymous referee for useful suggestions 
and comments on a previous version of this paper.



\begin{thebibliography}{00}

\bibitem[\protect\citeauthoryear{Alencar et al.}{2010}]{alencar10}
    Alencar, S. H. P., Teixeira P. S., Guimaraes M. M., McGinnis P. T., Gameiro J. F., et al., 2010, A\&A, 519, 88

\bibitem[\protect\citeauthoryear{Anders \& Grevesse}{1989}]{anders89}    
    Anders E., Grevesse N., 1989, GeCoA, 53, 197

\bibitem[\protect\citeauthoryear{Appenzeller et al.}{1983}]{appenzeller83}    
    Appenzeller I., Jankovics I., Krautter J., 1983, A\&AS, 53, 291

\bibitem[\protect\citeauthoryear{Bailey et al.}{2012}]{bailey12}    
    Bailey J. I., White R. J., Blake C. H., Charbonneau D., Barman T. S., Tanner A. M., Torres G., 2012, ApJ, 749, 16

\bibitem[\protect\citeauthoryear{Batalha \& Basri}{1993}]{batalha93}    
    Batalha C. C., Basri G., 1993, ApJ, 412, 363

\bibitem[\protect\citeauthoryear{Batalha et al.}{1998}]{batalha98}    
    Batalha C. C., Quast G. R., Torres C. A. O., Pereira P. C. R., Terra M. A. O., et al., 1998, 
    A\&A Supl. Ser., 128, 561


\bibitem[\protect\citeauthoryear{Claret et al.}{1995}]{claret95}
   Claret A., Diaz-Cordoves J., Gimenez A., 1995, A\&AS, 114, 247

\bibitem[\protect\citeauthoryear{Cody et al.}{2013}]{cody13}
    Cody A. M., Tayar J., Hillenbrand L. A., Matthews J. M., 
    Kallinger T., 2013, AJ, 145, 79

\bibitem[\protect\citeauthoryear{Cody et al.}{2014}]{cody14}
    Cody A. M., Stauffer J., Baglin A., Micela G., 
    Rebull L. M. et al., 2014, AJ, 147, 82

\bibitem[\protect\citeauthoryear{Comer{\'o}n}{2008}]{comeron08}    
    Comer{\'o}n F., 2008, Handbook of Star Forming Regions Vol. II, 
    Astronomical Society of the Pacific

\bibitem[\protect\citeauthoryear{Crockett et al.}{2012}]{crockett12}
    Crockett C. J., Mahmud N. I., Prato L., Johns-Krull C. M., 
    Jaffe D. T., Hartigan P. M., Beichman C. A., 2012, ApJ, 761, 164

\bibitem[\protect\citeauthoryear{Dekker et al.}{2004}]{Dekker04}
   Dekker H., D'Odorico S., Kaufer A., Delabre B., Kotzlowski H., 2004, SPIE, 4008, 534

\bibitem[\protect\citeauthoryear{Diaz-Cordoves et al.}{1995}]{diaz-cordoves95}
    Diaz-Cordoves J., Claret A., Gimenez A., 1995, A\&AS, 110, 329

\bibitem[\protect\citeauthoryear{Donati \& Collier Cameron}{1997}]{DC1997}
    Donati, J.-F., \& Collier Cameron, A. 1997, MNRAS, 291, 1

\bibitem[\protect\citeauthoryear{Donati et al.}{1997}]{Donati1997}
   Donati, J.-F., Semel, M., Carter, B.D., Rees, D.E.,
   Collier Cameron, A. 1997, MNRAS, 291, 658

\bibitem[\protect\citeauthoryear{Donati et al.}{2011}]{donati11}
     Donati J.-F., Gregory S.G., Alencar S.H.P., Bouvier J., 
     Hussain G., et al., 2011, MNRAS, 417, 472

\bibitem[\protect\citeauthoryear{Donati et al.}{2015}]{donati15}
     Donati J.-F., H{\'e}brard E., Hussain G. A. J., Moutou C., 
     Malo L., et al., 2015, MNRAS, 453, 3706

\bibitem[\protect\citeauthoryear{Errico et al.}{1996}]{errico96}    
    Errico L., Giovannelli F., Vittone A. A., 1996, MmSAI, 67, 209


\bibitem[\protect\citeauthoryear{Finkenzeller \& Basri}{1987}]{finkenzeller87}
    Finkenzeller U., Basri G., 1987, ApJ, 318, 823 

\bibitem[\protect\citeauthoryear{Fukugita et al.}{1996}]{fukugita96}
    Fukugita M., Ichikawa T., Gunn J. E., Doi M., Shimasaku K., Schneider D. P., 1996, AJ, 111, 1748

\bibitem[\protect\citeauthoryear{Gahm et al.}{1974}]{gahm74}
    Gahm G. F., Nordh H. L., Olofsson S. G., Carlborg N. C. J., 1974, A\&A, 33, 399

\bibitem[\protect\citeauthoryear{Gahm et al.}{1979}]{gahm79}
    Gahm G. F., Fredga K., Liseau R., Dravins D., 1979, A\&A, 73, L4


\bibitem[\protect\citeauthoryear{Gahm et al.}{2013}]{gahm13}
    Gahm G. F., Stempels H. C., Walter F. M., Petrov P. P., Herczeg G. J., 2013, A\&A, 560, 57





\bibitem[\protect\citeauthoryear{Giovannelli et al.}{1991}]{giovannelli91}
    Giovannelli F., Errico L., Vittone A. A., Rossi C., 1991, A\&AS, 87, 89

\bibitem[\protect\citeauthoryear{Giovannelli et al.}{1995}]{giovannelli95}
    Giovannelli F., Vittone A. A., Rossi C., Errico L., Bisnovatyi-Kogan G. S., 
    et al., 1995, A\&ASS, 114, 341

\bibitem[\protect\citeauthoryear{G{\"u}nther et al.}{2010}]{gunther10}
    G{\"u}nther H. M., Matt S. P., Schmitt J. H. M. M., G{\"u}del M., Li Z.-Y., Burton D. M., 2010, A\&A, 519, 97

\bibitem[\protect\citeauthoryear{Gray}{2010}]{gray10}
    Gray R. O., 2010, http://www.appstate.edu/~grayro/spec- trum/spectrum276/

\bibitem[\protect\citeauthoryear{Herbst et al.}{1994}]{herbst94}   
    Herbst W., Herbst D. K., Grossman E. J., Weinstein D., 1994, AJ, 108, 1906


\bibitem[\protect\citeauthoryear{Herczeg et al.}{2005}]{herczeg05}   
    Herczeg G. J., Walter F. M., Linsky J. L., Gahm G. F., Ardila  D. R., et al., 2005, AJ, 129, 2777

\bibitem[\protect\citeauthoryear{Hoffmeister}{1958}]{Hoffmeister58}
    Hoffmeister C., 1958, Veroff. Sternw. Sonnberg, 3, 333

\bibitem[\protect\citeauthoryear{Hoffmeister}{1965}]{Hoffmeister65}
    Hoffmeister C., 1965, Veroff. Sternw. Sonnberg, 6, 93

\bibitem[\protect\citeauthoryear{Hutchinson et al.}{1989}]{hutchinson89}
    Hutchinson M. G., Evans A., Davies J. K., Bode M. F., 1989, MNRAS, 237, 683

\bibitem[\protect\citeauthoryear{Joy}{1945}]{joy45}
    Joy A. H., 1945, ApJ, 102, 168

\bibitem[\protect\citeauthoryear{Juh{\'a}sz et al.}{2015}]{juhasz15}
    Juh{\'a}sz A., Benisty M., Pohl A., Dullemond C. P., Dominik C., Paardekooper S.-J., 
    2015, MNRAS, 451, 1147 

\bibitem[\protect\citeauthoryear{K{\"o}nigl}{1991}]{konigl91}
    K{\"o}nigl A., 1991, ApJ, 370, L39

\bibitem[\protect\citeauthoryear{Kulkarni \& Romanova}{2008}]{kulkarni08}  
     Kulkarni A. K., Romanova M. M., 2008, MNRAS, 386, 673 

\bibitem[\protect\citeauthoryear{Kulkarni \& Romanova}{2009}]{kulkarni09}  
     Kulkarni A. K., Romanova M. M., 2009, MNRAS, 398, 701 

\bibitem[\protect\citeauthoryear{Kulkarni \& Romanova}{2013}]{kulkarni13}  
     Kulkarni A. K., Romanova M. M., 2013, MNRAS, 433, 3048 

\bibitem[\protect\citeauthoryear{Kurosawa \& Romanova}{2013}]{kurosawa13}
    Kurosawa R., Romanova M. M., 2013, MNRAS, 431, 2673

\bibitem[\protect\citeauthoryear{Kurucz}{1992}]{kurucz92}
    Kurucz R., 1992, IAUS, 149, 225


\bibitem[\protect\citeauthoryear{Lagrange et al.}{2013}]{lagrange13}    
    Lagrange A.-M., Meunier N., Chauvin G., Sterzik M., Galland F., Lo Curto G., 
    Rameau J., Sosnowska D., 2013, A\&A, 559, A83

\bibitem[\protect\citeauthoryear{Lamzin et al.}{1996}]{lamzin96}
    Lamzin S. A., Bisnovatyi-Kogan G.S., Errico L., Giovannelli F., Katysheva N. A., 
    Rossi C., Vittone A. A., 1996, A\&A, 306, 877    

\bibitem[\protect\citeauthoryear{Matthews et al.}{2004}]{M2004}
    Matthews J. M., Kusching R., Guenther D. B., Walker G. A. H.,
    Moffat A. F. J., Rucinski S. M., Sasselov D., Weiss W. W., 
    2004, Nature, 430, 51

\bibitem[\protect\citeauthoryear{Martin et al.}{1994}]{martin94}
   Martin E. L., Rebolo R., Magazzu A., Pavlenko Y. V., 1994, A\&A, 282, 503

\bibitem[\protect\citeauthoryear{Mawet et al.}{2012}]{mawet12}
   Mawet D., Absil O., Riaud P., Surdej J., Montagnier G., et al., 2012, A\&A, 544, 131

\bibitem[\protect\citeauthoryear{Mayor et al.}{2003}]{Mayor03}
   Mayor M., Pepe F., Queloz D., Bouchy F., Rupprecht G., et al., 2003, The Messenger, 114, 20
 
\bibitem[\protect\citeauthoryear{Menzies et al.}{1989}]{menzies89}
   Menzies J. W., Cousins A. W. J., Banfield R. M., Laing J. D., 1989, {\it SAAO} Circulars, 13, 1-13
   
\bibitem[\protect\citeauthoryear{Merrill}{1941}]{Merrill41}
   Merrill P. W., 1941, PASP, 53, 342

\bibitem[\protect\citeauthoryear{Motl}{2011}]{Motl11}
   Motl D., 2011, http://c-munipack.sourceforge.net

\bibitem[\protect\citeauthoryear{Pani{\'c} et al.}{2009}]{panic09}
   Pani{\'c} O., Hogerheijde M. R., Wilner D., Qi C., 2009, A\&A, 501, 269


\bibitem[\protect\citeauthoryear{Pinte et al.}{2008}]{pinte08}
   Pinte C., Padgett D. L., Menard F., Stapelfeldt K. R., Schneider G., et al., 2008, A\&A 489, 633

\bibitem[\protect\citeauthoryear{Pojmanski \& Maciejewski}{2004}]{pojmanski04}
   Pojmanski, G., Maciejewski, G., 2004, AcA, 54, 153

\bibitem[\protect\citeauthoryear{Pribulla \& Rucinski}{2008}]{pribulla08}
   Pribulla T., Rucinski S. M., 2008, MNRAS, 386, 377


\bibitem[\protect\citeauthoryear{Plagemann}{1969}]{Plagemann69}
   Plagemann S., 1969, Mem. Soc. Roy. Sci. Liege, 5th Series, 19, 331

\bibitem[\protect\citeauthoryear{Reipurth et al.}{1996}]{reipurth96}    
     Reipurth Bo, Pedrosa A., Lago M. T. V. T., 1996, A\&AS, 120, 229

\bibitem[\protect\citeauthoryear{Romanova et al.}{2004}]{romanowa04}    
     Romanova M. M., Ustyugova G. V., Koldoba A. V., Lovelace R. V. E., 2004, ApJ, 610, 920


\bibitem[\protect\citeauthoryear{Romanova et al.}{2008}]{romanowa08}    
     Romanova M. M., Kulkarni A. K., Lovelace R. V. E., 2008, ApJ, 673, L171

\bibitem[\protect\citeauthoryear{Romanova \& Kulkarni}{2009}]{romanowa09}    
    Romanova M. M., Kulkarni A. K., 2009, MNRAS, 398, 1105 

\bibitem[\protect\citeauthoryear{Rucinski}{1992}]{ruc92}
    	Rucinski S. M., 1992, AJ, 104, 1968

\bibitem[\protect\citeauthoryear{Rucinski}{2002}]{ruc02}
    	Rucinski S. M., 2002, AJ, 124, 1746

\bibitem[\protect\citeauthoryear{Rucinski et al.}{2008}]{ruc08}
    	Rucinski S. M., Matthews J. M., Kuschnig R., Pojmanski G., 
        Rowe J., et al., 2008 MNRAS, 391, 1913

\bibitem[\protect\citeauthoryear{Rucinski et al.}{2010}]{ruc10}
    Rucinski S. M., Zwintz K., Hareter M., Pojmanski G., Kuschnig R., 
    et al., 2010, A\&A, 522, 113

\bibitem[\protect\citeauthoryear{Rucinski}{2010}]{ruc2010}        
    Rucinski, S. M. 2010,
    ASP Conf., 435, 195

\bibitem[\protect\citeauthoryear{Rucinski}{2012}]{ruc2012}        
    Rucinski, S. M. 2012, in 
    From Interacting Binaries to Exoplanets: 
    Essential Modeling Tools, IAU Symp. 282,
    M. Richards and I. Hubeny (eds), p.365

\bibitem[\protect\citeauthoryear{Rucinski et al.}{2013}]{ruc13}
    Rucinski S. M., Pribulla T., Budaj J., 2013, AJ, 146, 70

\bibitem[\protect\citeauthoryear{Rucinski}{2015}]{ruc15}
    Rucinski S. M., 2015, AJ, 149, 49

\bibitem[\protect\citeauthoryear{Schwartz \& Heuermann}{1981}]{schwartz81}
    Schwartz R. D., Heuermann R.,1981, AJ, 86, 1526
    

\bibitem[\protect\citeauthoryear{Siwak et al.}{2010}]{siwak10}
    Siwak M., Zola S., Koziel-Wierzbowska D., 2010, AcA, 60, 305

\bibitem[\protect\citeauthoryear{Siwak et al.}{2011a}]{siwak11a}
    Siwak M., Rucinski S. M., Matthews J. M., Pojmanski G., 
    Kuschnig R., et al., 2011a, MNRAS, 410, 2725

\bibitem[\protect\citeauthoryear{Siwak et al.}{2011b}]{siwak11b}
    Siwak M., Rucinski S. M., Matthews J. M., Kuschnig R., Guenther D. B., et al., 
    2011b, MNRAS, 415, 1119 

\bibitem[\protect\citeauthoryear{Siwak et al.}{2013}]{siwak13}
    Siwak M., Rucinski S. M., Matthews J. M., Kuschnig R., Guenther D. B., et al., 2013, MNRAS, 432, 194

\bibitem[\protect\citeauthoryear{Siwak et al.}{2014}]{siwak14}
    Siwak M., Rucinski S. M., Matthews J. M., Kuschnig R., Guenther D. B., et al., 2014, MNRAS, 444, 327

\bibitem[\protect\citeauthoryear{Stauffer et al.}{2014}]{stauffer14}
    Stauffer J., Cody A. M., Baglin A., Alencar S., Rebull L., et al., 2014, AJ, 147, 83

\bibitem[\protect\citeauthoryear{Stempels \& Piskunov}{2002}]{stempels02}
    Stempels H. C., Piskunov N., 2002, A\&A 391, 595

\bibitem[\protect\citeauthoryear{Stempels et al.}{2007}]{stempels07}
    Stempels H. C., Gahm G. F., Petrov P. P.,2007, A\&A 461, 253, 259

\bibitem[\protect\citeauthoryear{Stetson}{1987}]{stet}
    Stetson P. B., 1987 PASP, 99, 191

\bibitem[\protect\citeauthoryear{Strassmeier et al.}{2005}]{strassmeier05}
    Strassmeier K. G., Rice J. B., Ritter A., K{\"u}ker M., Hussain G. A. J., 
    Hubrig S., Shobbrook R., 2005, A\&A, 440, 1105


\bibitem[\protect\citeauthoryear{Walker et al.}{2003}]{WM2003}
    Walker G. A. H., Matthews J. M., Kuschnig R., Johnson R., Rucinski S. M., et al., 
    2003, PASP, 115, 1023

\bibitem[\protect\citeauthoryear{Warmels}{1991}]{Warmels91}
    Warmels R. H., 1991, PASP Conf. Series, 25, 115

\bibitem[\protect\citeauthoryear{Wichmann et al.}{1998}]{wichmann98}
    Wichmann R., Bastian U., Krautter J, Jankovics I., Rucinski S. M., 1998, MNRAS, 301, L39

\bibitem[\protect\citeauthoryear{Wichmann et al.}{1999}]{wichmann99}
    Wichmann R., Covino E., Alcala J. M., Krautter J. Allain S., Hauschildt P. H., 1999, MNRAS, 307, 909

\bibitem[\protect\citeauthoryear{Windemuth et al.}{2013}]{windemuth13}
    Windemuth D., Herbst W., Tingle E., Fuechsl R., Kilgard R., Pinette M., Templeton M., 
    Henden A., 2013, ApJ, 768, 67

\bibitem[\protect\citeauthoryear{Wilson}{1979}]{wilson79}
    Wilson R. E., 1979, ApJ, 234, 1054

\bibitem[\protect\citeauthoryear{Valenti et al.}{2003}]{valenti03}
   Valenti J. A., Fallon A. A., Johns-Krull C. M., 2003, AJSS, 147, 305

\bibitem[\protect\citeauthoryear{van Kempen et al.}{2007}]{vanKempen07}
    van Kempen T. A., van Dishoeck E. F., Brinch C., Hogerheijde M. R., 2007, A\&A 461, 983


\bibitem[\protect\citeauthoryear{Zwintz et al.}{2013}]{zwintz13}
    Zwintz K., Fossati L., Guenther D. B., Ryabchikova T., Baglin, A., et al., 2013, A\&A, 552A, 68

\end{thebibliography}
\end{document}